\documentclass[twocolumn]{aastex62}
\usepackage[utf8]{inputenc}
\usepackage{graphicx}
\usepackage{natbib}
\pdfoutput=1

\newcommand{\civ}{\ion{C}{4}}
\newcommand{\ciii}{\ion{C}{3}]}
\newcommand{\mgii}{\ion{Mg}{2}}
\newcommand{\lledd}{$L/L_{\rm Edd}$}

\begin{document}

\title{Characterizing Quasar CIV Emission-line Measurements from Time-resolved Spectroscopy}

\author{Angelica B. Rivera*}
\affil{Department of Physics, 32 S.\ 32nd Street, Drexel University, Philadelphia, PA 19104}

\author{Gordon T. Richards*}
\affil{Department of Physics, 32 S.\ 32nd Street, Drexel University, Philadelphia, PA 19104}
\email{*Email: abr54@drexel.edu (ABR); gtr@physics.drexel.edu (GTR)}

\author{Paul C. Hewett}
\affil{Institute of Astronomy, University of Cambridge, Madingley Road, Cambridge CB3 0HA}

\author{Amy L. Rankine}
\affil{Institute of Astronomy, University of Cambridge, Madingley Road, Cambridge CB3 0HA}

\received{2020 June 19}

\accepted{2020 July 13}

\begin{abstract}

We use multi-epoch quasar spectroscopy to determine how accurately single-epoch spectroscopy can locate quasars in emission-line parameter space in order to inform investigations where time-resolved spectroscopy is not available.  We explore the improvements in emission-line characterization that result from using non-parametric information from many lines as opposed to a small number of parameters for a single line, utilizing reconstructions based on an independent component analysis applied to the data from the Sloan Digital Sky Survey Reverberation Mapping project.
We find that 
most of the quasars are well described by just two components, while more components signal a quasar likely to yield a successful reverberation mapping analysis. 
In single-epoch spectroscopy the apparent variability of equivalent width is exaggerated because it is dependent on the continuum. 
Multi-epoch spectroscopy 
reveals that single-epoch results do not significantly change where quasars are located in \civ\ parameter space and do not have a significant impact on investigations of the global Baldwin Effect.  
Quasars with emission line properties indicative of higher \lledd\ are less variable, 
consistent with models with enhanced accretion disk density.
Narrow absorption features at the systemic redshift may be indicative of orientation (including radio-quiet quasars) and may appear in as much as 20\% of the quasar sample.  Future work applying these techniques to lower luminosity quasars will be important for understanding the nature of accretion disk winds.

\end{abstract}


\vspace{0.5cm}

\section{Introduction}

To understand the physics of quasars and active galactic nuclei (AGN) more generally, it is ultimately necessary to know the key physical parameters for individual objects.  It is thought that the driving physical parameters are mass, accretion rate, and spin of the central black hole \citep[BH; e.g.,][]{Netzer2013}, where the inherent asymmetry of accretion disk models means that orientation must also play a role in how individual quasars manifest---even for unobscured systems \citep[e.g.,][]{Shen+2014}.

It is generally difficult to do more than estimate the BH mass and accretion rate (using BH mass ``scaling relations", e.g., \citealt{Krolik2001,Vestergaard+2006,Coatman+2017}), and it is nearly impossible to determine accurate orientation and spin parameters for large numbers  of individual objects.  As such, investigations of quasars are constrained in a way that limits physical interpretation. We nevertheless have access to {\em empirical} parameter spaces that must encode these very physical parameters.  While there is no sign that the quasar population---at least as captured by existing spectroscopy from the Sloan Digital Sky Survey (SDSS; \citealt{York+2000})---is anything but continuous, it is nevertheless possible to isolate subsets in such empirical parameter spaces (e.g., \citealt{Sulentic+2007,Richards+2011,Rankine+2020}).  Unless there exist significant degeneracies between physical parameters, extrema in empirical parameter spaces can be used to isolate extrema in physical parameter spaces---without actually knowing the physical parameters.

The most well known of these empirical parameter spaces is the so-called ``Eigenvector 1" (EV1) parameter space, the components of which are summarized by \citet{Brotherton+1999}.  At low-redshift the EV1 parameter space
can be visualized by using FWHM H$\beta$ versus R(FeII) \citep{Boroson+1992,Shen+2014}.  Here R(FeII) is defined as the ratio of the equivalent width (EQW) of the FeII emission-line complex covering between 4344--4684$\textrm{\AA}$ and the EQW of broad H$\beta$; R(FeII) is believed to be correlated with the Eddington ratio, \lledd.  The FWHM H$\beta$ is believed to be an indicator of some combination of orientation and black hole mass \citep{Shen+2014}.  Principal component analysis (PCA) of these EV1 parameters \citep{Boroson+1992} finds that most of the statistical variance (or here, sub-classification diversity) is described by two eigenvectors, which \citet{Boroson2002} argue are tracers for \lledd\ and $\dot{M}$, respectively.

At high-redshift the empirical variance of quasars instead is best captured by the \civ\ $\lambda$1549\AA\ emission line
(e.g., \citealt{Wills+1999,Sulentic+2007}).  It has been shown that two properties of the \civ\ emission line,  EQW and blueshift \citep{Gaskell1982}, capture most of the diversity of the broad emission-line properties of high-redshift quasars (e.g., \citealt{Sulentic+2007,Richards+2011}). It has been argued (e.g, \citealt{Richards+2011,Wang+2012}) that the changes in these two properties reflect a two-component (disk-wind) model of the broad-emission-line region (BELR; e.g., \citealt{Collin+2006}).  In such a model, UV photons drive the gas through radiation-line pressure, and X-rays work to counteract the process, stripping the gas of electrons before the UV photons are able to drive a wind (e.g., \citealt{Murray+1995,Proga+2000}). As summarized from a theoretical perspective by \citet{Giustini+2019}, the \civ\ BELR can be thought of as dominated by a wind-driven component at one end of the \civ\ distribution (high-blueshift, low EQW) and a disk (or perhaps a failed wind) at the other (high EQW, low-blueshift).  BH mass and \lledd\ drive the main observed differences---by regulating how much volume of the accretion disk ``atmosphere" participates successfully in wind driving.
For a given \lledd\ the BH mass controls the temperature of the accretion disk and, as a result, the ability to generate radiation line-driven winds \citep{Giustini+2019}.
While there is no one-to-one agreement between \civ\ properties and \lledd, the mean \lledd\ increases with increasing blueshift and decreasing EQW \citep{Shemmer+2015,Rankine+2020}. Thus for high-redshift quasars, spectroscopy of even just the \civ\ emission line region may be a powerful empirical probe of BH mass and accretion rate, even in the absence of accurate estimates of those parameters. 

However, survey-quality, single-epoch spectroscopy yields noisy measurements of the \civ\ emission-line parameters.    Moreover, variability of both the continuum and the BELR leads to further uncertainty with regard to the ``true" location of any quasar in the \civ\ parameter space.  In this investigation we seek to determine the true distribution of quasars in \civ\ parameter space and how variability (in addition to limited spectrum S/N) may cause the location of an individual object to change with time (thus adding scatter to the distribution as measured from single-epoch spectra).  The SDSS Reverberation Mapping project (SDSS-RM; \citealt{SDSSRM}) provides an ideal laboratory for this work. \citet{Sun+2018} used this same sample to explore the question of whether the \civ\ emission line blueshift is variable, while \citet{Dyer+19} used this sample to explore the Baldwin (\citeyear{Baldwin1977}) Effect and bluer-when-brighter \citep{Giveon+99,Schmidt+12} correlations.  Herein we extend those investigations using a novel analysis technique that helps to extract maximal information from the data.

Two aspects of this data and our analysis enable minimization of sources of error/scatter (including line variability and continuum variability). First, stacking of many individual epochs results in a very high S/N median spectrum for each object. Second, fitting reconstructed spectra based on the application of independent component analysis (ICA; \citealt{Allen+2013}) allows leveraging correlated information from other UV emission lines (particularly, \ion{He}{2} $\lambda$1640\AA, \ion{Al}{3} $\lambda$1857\AA, \ion{Si}{3}] $\lambda$1892\AA, \ion{C}{3
}] $\lambda$1909\AA, and \ion{Mg}{2} $\lambda$2799\AA) to more accurately measure the \civ\ line properties from individual (and necessarily more noisy) epochs. ICA-reconstructed spectra (\S~\ref{sec:ica}) significantly reduce the scatter in emission-line parameter measurements revealing less noisy trends with time.

With a more robust understanding of the \civ\ space, we are able to draw conclusions about what variability can reveal about the physics of the BELR (e.g., in the context of the \citealt{Giustini+2019} model) and also about the nature of the well-known Baldwin Effect (e.g., \citealt{Baldwin1977, Dietrich+2002, Baskin+2004, Shemmer+2015}), which connects the \civ\ EQW to the luminosity (or perhaps \lledd).
Furthermore, knowing how the \civ\ emission-line properties vary within the \civ\ parameter space is a necessary condition for being able to explore how the emission-line properties change with luminosity (including exploring biases in the properties of classical RM samples as reported by \citealt{Richards+2011}) and how the emission line properties change with X-ray properties---as expected theoretically, \citep[e.g.,][]{Giustini+2019}, and as seen empirically (e.g., \citealt{Kruczek+2011,Timlin+2020}).  We will probe to lower luminosity and add X-ray analyses to the work herein in 
Rivera (2020, in preparation).

The paper is organized as follows. Sections~\ref{sec:data} and \ref{sec:ica} describe the data sets used in this analysis, and the types of ICA used to build the reconstructed spectra, respectively. In Section \ref{sec:lineparams} we describe the line parameters we chose and our methods of extracting them. In Section \ref{sec:disc} we discuss the effects of variability on quasars within the \civ\ parameter space, as well as trends (or lack thereof) of interesting classes of quasars within that space. Our conclusions are given in Section \ref{sec:conclusion}. 

\section{Data}
\label{sec:data}

\subsection{SDSSRM Sample}
\label{sec:sample}

The multi-epoch spectra from the SDSS-RM program \citep{SDSSRM} provide an ideal laboratory for simultaneously investigating the diversity of BELR properties in quasars and how that diversity is affected by both measurement error and variability.  The data set consists of years of optical spectroscopy taken with the BOSS (Baryon Oscillation Spectroscopic Survey; \citealt{BOSS}) spectrograph as part of the SDSS-III \citep{SDSSIII} project. Spectra were recorded for 849 broad-line quasars in the area of a single $\simeq$7 $\textrm{deg}^{2}$ SDSS plate.  The spectra cover 3650--10,400\AA\ at a resolution of $R \sim 2000$, with S/N such that an object with $g=21.2$ has S/N~$>4.5$ (as compared to 3.9 for other SDSS-III programs). In addition to higher S/N, the SDSS-RM program used 3.5 times as many spectrophotometric standards in attempt to achieve better spectrophotometric calibration than SDSS-III as a whole \citep{Margala+16}. Further supporting this calibration was ground-based $g$- and $i$-band photometry taken roughly every 2 days (as compared to the spectroscopic cadence of $\simeq 4$ days in the first year).

\begin{figure}[th!]
    \includegraphics[width=3.5in]{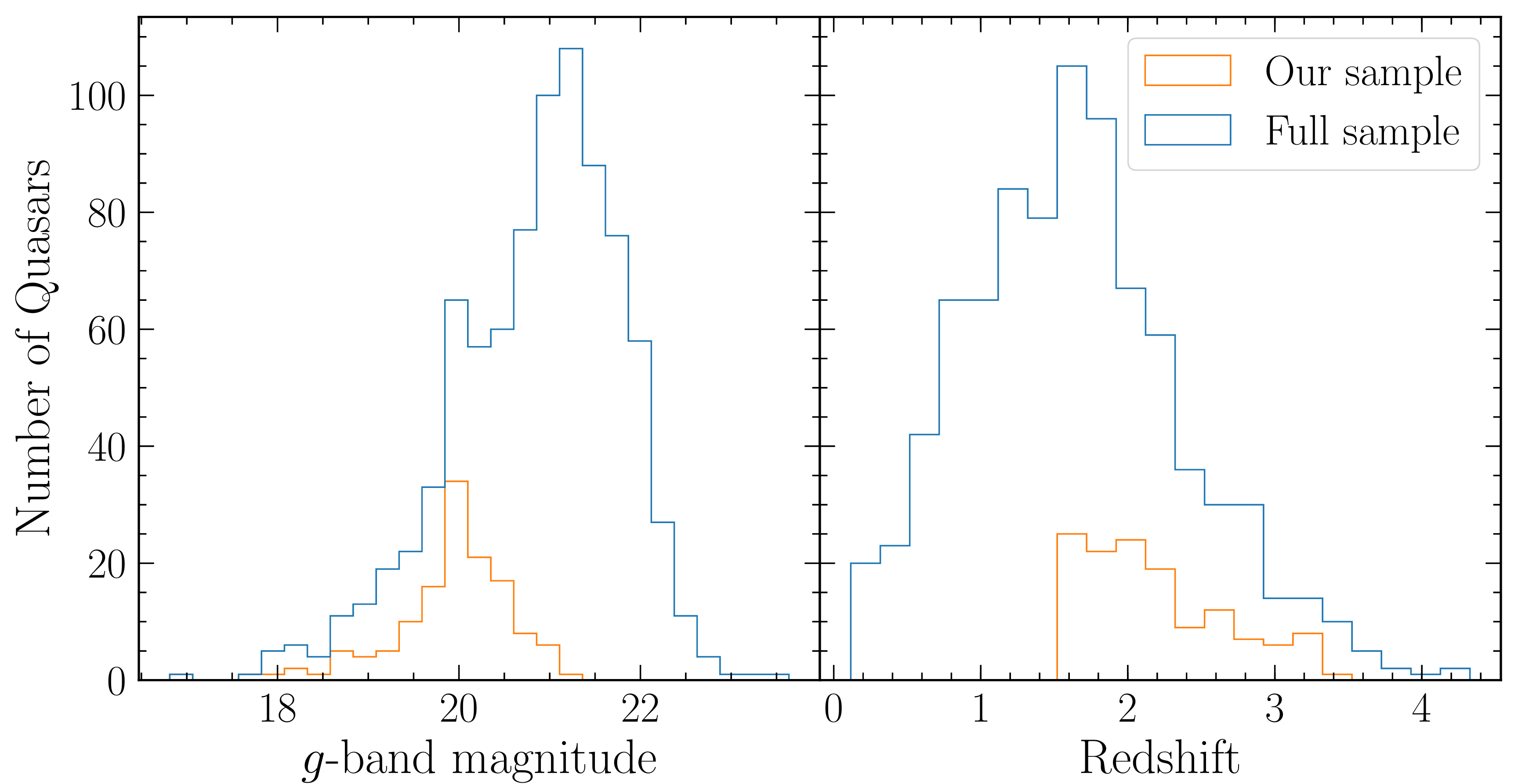}
    \caption{$g$-band magnitudes and redshifts for the quasars in our final sample.   Shown are both the full SDSS-RM sample (blue) and the high-redshift, high S/N subsample analyzed herein (orange).  For comparison, the analyses of \citet{Sun+2018} and \citet{Dyer+19} used 362 quasars over $1.48<z<2.6$ and 340 quasars over $1.62<z<3.3$, respectively.
    \label{fig:fig1}}
\end{figure}

A total of 556 objects from this sample possess the \ion{C}{4} $\lambda$1549$\textrm{\AA}$ line, 552 of which were publicly released in SDSS DR14. Out of those 552, 133 had 30 or more epochs with spectra possessing a mean S/N ($>$ 6) over 4500-8000\,\AA \ allowing us to perform the analysis as described below. The redshift range of the quasars is $1.57 \le z \le 3.39$. In our analysis we utilize the first 53 epochs taken over a span of three years. The first 32 were taken between January and July of 2014 with an average cadence of four days. For the next two years the cadence was twice a month over the same range of months. As a result of the cut in S/N, the epochs included will not always be the same for each object, nor will they fully represent this cadence. This analysis does not include the last 3 epochs of Season 3 (from 2016). We removed epochs 3 and 7 from the sample, as they both had low S/N.  Future work on the full sample would be able to take advantage of 67 epochs covering 4 seasons from 2014 through 2017, with additional epochs in 2018 and 2019 \citep{Shen+2019}. The distribution of $g $-magnitude and redshift for both our objects and the full SDSS-RM sample is shown in Figure~\ref{fig:fig1}. The faintest of these had $g$-band magnitudes of $\simeq$ 21. 

\citet{Grier+2019} computed \civ\ lag times for a subsample of SDSS-RM quasars on the way to estimating BH masses for the quasars.  For 18 objects \citet{Grier+2019} found these lags to be robust (rough agreement between 3 analyses), defining a ``gold sample," with lags between $\sim$60 days to $\sim$450 days in the observed frame. Given the redshifts of the quasars, the average rest-frame time lag for objects within the gold sample is $\sim$70 days. Each observing season is about 180 days (60 days in the rest frame at $z=2$), with the full time covered by the 53 epochs being 910 days (300 days in the rest frame at $z=2$, where, given the observing seasons, only about half that time is sampled). Four of the gold sample quasars are within our sample (RMIDs 275, 298, 387, 401). There are additionally 11 objects from our full sample with lag estimates in their Table 3. These are RMIDs 36, 128, 130, 201, 231, 237, 389, 408, 485, 554, 562. Therefore the majority of our sample lack well-defined lag times.  By design the SDSS-RM program has a cadence sufficient for the bulk cross-correlation of continuum and broad line region light curves needed for RM analysis \citep{SDSSRM}; however, due to the ground-based (seasonal) nature of the program and the long time lags of luminous quasars, even those quasars with known time lags lack the time coverage needed to provide simultaneous measurements of the rest-frame continuum and BELR.  For example a quasar with an observed-frame lag of 180 days will have essentially no simultaneous observations of the continuum and the BELR in the rest frame, whereas a quasar with an observed-frame time lag of 360 days will have BELR data in years 2 and 3 that correspond with continuum data from years 1 and 2.  We discuss how RM signatures may manifest in spectral features in Section \ref{sec:individualica}, how we account for the lag times in our \civ\ parameter measurements in Section~\ref{sec:eqw}.  

\subsection{SDSS-RM Data Processing}
\label{sec:dataproc1}

The spectra for the SDSS-RM program were processed in multiple ways and it is important to recognize the differences between those processes.  Most of our analysis was performed on the standard ``BOSS" spectral reductions.  This choice will allow analysis of thousands of SDSS-BOSS quasars (not just the SDSS-RM objects) using the same methods and characterization of their locations within the \civ\ parameter space.

Additional steps were taken by the SDSS-RM program to improve the absolute spectrophotometry of the SDSS-RM spectra.  First are the wavelength-dependent corrections to the spectral reductions as illustrated in Figure~18 of \citet{SDSSRM}.  These ``Custom" reductions improve upon the BOSS reductions by considering the wavelength-dependent light losses across the plate.  As the plates are drilled for a particular time and the observations are long, it is expected that objects farther from the center of the plate will experience more light losses and that these losses will be wavelength dependent.  The Custom reductions characterize and attempt to correct for these losses using additional calibration stars (beyond the number used in the main BOSS program).  Second, are wavelength-independent ``PrepSpec" corrections described in Section~2.1 of \citet{PrepSpec} that are determined by assuming that the narrow lines are not variable.  These corrections further enhance the spectrophotometry by accounting for wavelength-independent light losses.  The combination of Custom reductions and PrepSpec corrections give the best possible absolute spectrophotometry for the SDSS-RM objects.  \citet{PrepSpec} discusses the expected level of uncertainty that remains.  In the analysis herein we will utilize different reductions/calibrations depending on the goal of the analysis.

\begin{figure*}[t]
\includegraphics[width=6.5in, trim= -2cm 0 0 0]{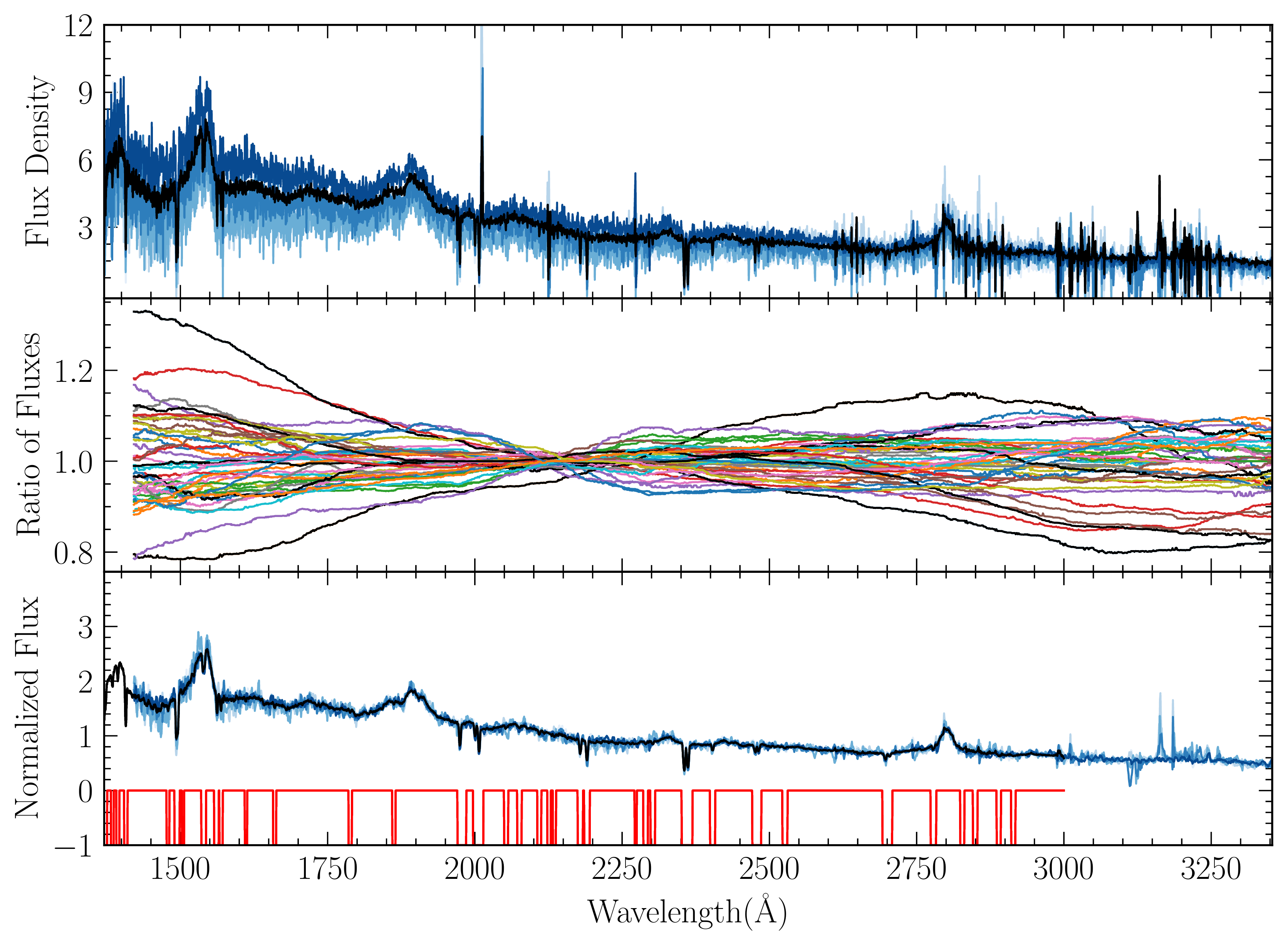}
\caption{Example of the procedure for correcting observation-induced spectrophotometric differences between epochs. The quasar is RMID 725 and epochs 16, 17, 30, 40, and 47 are shown. Top panel: the original spectra (shades of blue, lighter colors indicate earlier epochs) and the flux density of the median-composite of all 42 epochs (shown in black). The units for flux density in this paper are $10^{-17}$ erg/s/cm$^{2}$/\AA. Middle panel: the ratio-spectra for each epoch normalized at 2500\,\AA \ (epochs included in the top panel are shown in black). Bottom panel: overplot of the original spectra for the five epochs multiplied by their ratio-spectra and inverse-variance-weighting smoothed with a seven-pixel window. The black spectrum is the median spectrum upon which the global ICA reconstruction (\S~\ref{sec:globalica}) was performed. The mask used to eliminate absorption features from the spectra is shown in red.
\label{fig:fig2}}
\end{figure*}

\subsection{Preprocessing for ICA Reconstruction}
\label{sec:dataproc2}

While the SDSS-RM spectra have undergone a careful spectrophotometric calibration process, our analysis is focused on examining the intrinsic variability of the quasars in the sample.  Our primary analysis uses the BOSS reductions; as such, we needed to first preprocess the data to eliminate spectral shape changes due to any residual observational effects.  For each unique quasar, we first compute a median spectrum then ``morph" each of the individual epochs to have the same spectral shape as the median.
This process effectively corrects for spectrophotometric differences due to observational effects (e.g., differences in airmass, variations due to the different times of observation, optical fibers slipping) that may not have been corrected fully as part of the BOSS reductions. The procedure is comparable to that described by \citet{Rankine+2020}, except that here we only morph each epoch internally (forcing them to match their median composite spectrum)---rather than morphing all of the objects to some common continuum that applies to all of the objects.  

In more detail, a median-composite spectrum was calculated using all the individual epochs. A search for narrow absorption systems in the composite spectrum is undertaken to generate a mask that flags pixels affected by absorber systems (see Section~4.2 of \citealt{Rankine+2020}). The median-composite spectrum was divided into each individual-epoch spectrum. At the start and finish wavelengths, the median-divided spectra were reflected and then inverted about the median value of the last 150 pixels.\footnote[3]{The procedure ensures that any large-scale trends present in the median-divided spectra are retained at the spectrum boundaries following the next filtering stage.} A 601-pixel median filter is then applied to the median-divided spectra. The result is a very smooth ``ratio-spectrum" that incorporates the large-scale wavelength-dependent differences in flux between the individual quasar spectrum and the median-composite spectrum. The original quasar spectrum is multiplied by the ratio-spectrum to produce a version with the same shape as the median-composite spectrum. Finally, the transformed spectra are smoothed using inverse-variance weighting over a seven-pixel window to increase the S/N, particularly longward of 7000\,\AA.  
    
The results of the procedure are shown for one quasar in Figure~\ref{fig:fig2}. The top panel shows five epochs from RMID 725, as well as the median-composite of all epochs (black solid line). The middle panel shows the ratio-spectra for all epochs with the five epochs shown in the top panel highlighted in black. The significant spectrophotometric differences between the different epochs are evident. The bottom panel shows the spectra of the five epochs following multiplication by the ratio-spectra together with the median-composite spectrum. We also show the mask array used to identify ``bad" pixels.

\section{Independent Component Analysis}
\label{sec:ica}

We seek to create a significantly higher S/N reconstruction of the SDSS-RM spectra from which to measure line parameters.  For this process we adopt independent component analysis (ICA; \citealt{ICA94,Hastie}), which is a blind source separation algorithm. For systems that can be described as the sum of independent parts (e.g., host galaxy, broad line region, narrow line region), blind source separation algorithms, like ICA, are potentially very powerful techniques.
The particular brand of ICA adopted here is mean-field ICA (MFICA; \citealt{Hojen-Sorenson2002Mean-FieldAnalysis,Allen+2013}).
Each spectrum is reconstructed from a basis set of ``components," with each individual spectrum characterized by the different ``weights" of the components.  While ICA is not intended to handle non-linear correlations in the data, careful application of ICA to smaller subsets of the data might be expected to overcome this constraint.
In practice we compute three different types of ICA reconstruction for each object in the SDSS-RM sample.  First is a ``global" ICA (\S~\ref{sec:globalica}) where the components are derived from the quasar population as a whole.  These global reconstructions can be generated for any single-epoch quasar spectra (such as BOSS quasars that are not in the SDSS-RM field).  Next is an ``individual" ICA (\S~\ref{sec:individualica}) where the components are derived from the individual spectroscopic epochs of a given quasar.  The individual reconstructions are specific to the multi-epoch nature of the SDSS-RM program.  Lastly, we combine the methods to produce absorption-corrected reconstructions at each epoch (\S~\ref{sec:CombRecon}).  In all cases, our use of ICA components is not meant to represent physically distinct inputs. We are merely utilizing the ICA components to create higher S/N reconstructions of the spectra.

\subsection{Global ICA} 
\label{sec:globalica}

Global ICA reconstructions were produced for the {\em median spectra} of the 133 SDSS-RM quasars.  The ICA components were taken from the analysis of \citet{Rankine+2020}, over a rest wavelength range of 1275-3000 $\textrm{\AA}$.  Following \citet{Rankine+2020} we split the sample into three groups: the whole sample, a high EQW subsample, and a low EQW subsample. The reconstructions required 10, 7, and 10 components, respectively.  The best of these three reconstructions (as determined by eye) were adopted as the final global reconstruction.  To improve the emission-line reconstruction of quasars exhibiting absorption (such as BALQSOs), the automatically-derived absorber-masks were visually inspected and the size of the absorber-mask adjusted as needed. Improved redshifts were generated prior to generating the ICA reconstructions using the same method as \citet{Rankine+2020}. The average change in redshift was 0.0011, corresponding to $\sim$110 km $\textrm{s}^{-1}$.

\subsection{Individual ICA}
\label{sec:individualica}

Since we are interested in quasar emission-line properties not just as a snapshot in time, but as a function of time, we also performed ICA analysis on the individual epochs of each quasar. The first such analysis of this type (dimensionality reduction on multiple epochs of the same quasar as opposed to single epochs of many quasars) was performed by \citet{Mittaz+90}.  Our ``individual ICA" analysis treats each of the epochs as a unique input, using ICA to capture the changes in the emission lines {\em of each quasar}.  The first component was defined to be the median spectrum. For emission lines where the changes are dominated by changes to the amplitude of the line, or by changes to the continuum level, we find that dimensionality reduction techniques like ICA require just one additional component.  If more complicated changes to the emission lines occur (e.g., both the amplitude and FWHM change), additional components are needed to encode the time-dependent shape of the emission lines.  

\begin{figure}[t!]
    \includegraphics[width=3.5in]{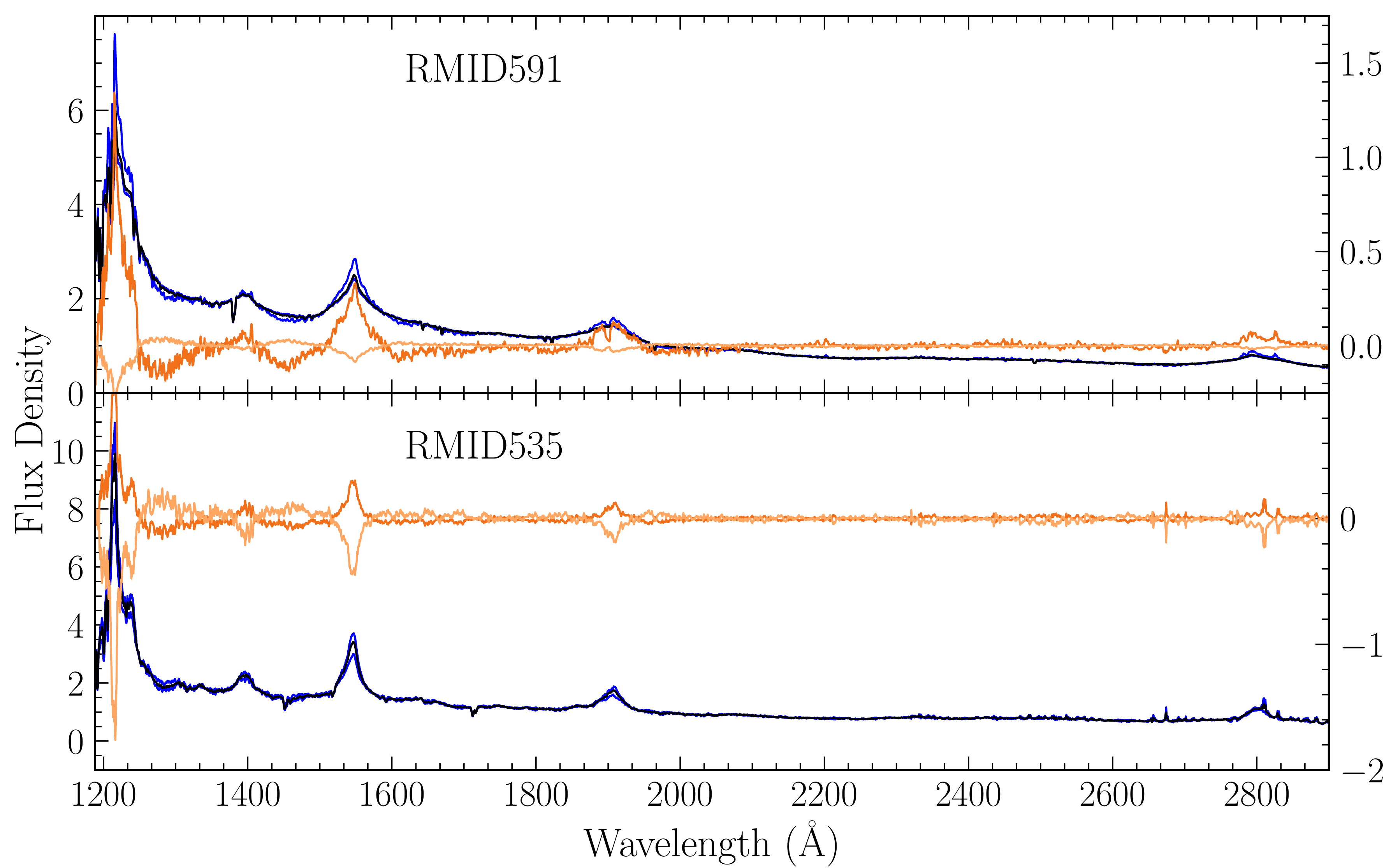}
    \caption{The median ICA reconstruction (dark blue) with the reconstruction with the smallest and largest W2 values overlaid (light blue) for two of the objects with the largest amount of variability in the sample. The smallest and largest W2*Component 2 are shown in gold. Both the left and right axis show the flux density of the spectra, the right axis shows the strength of W2*Component 2. Notice that the emission line changes are correlated across Ly$\alpha$, \ion{Si}{4}, \civ, \ion{C}{3}], and \ion{Mg}{2}. 
    \label{fig:fig3}}
\end{figure}

\begin{figure*}
    \includegraphics[width=6.5in]{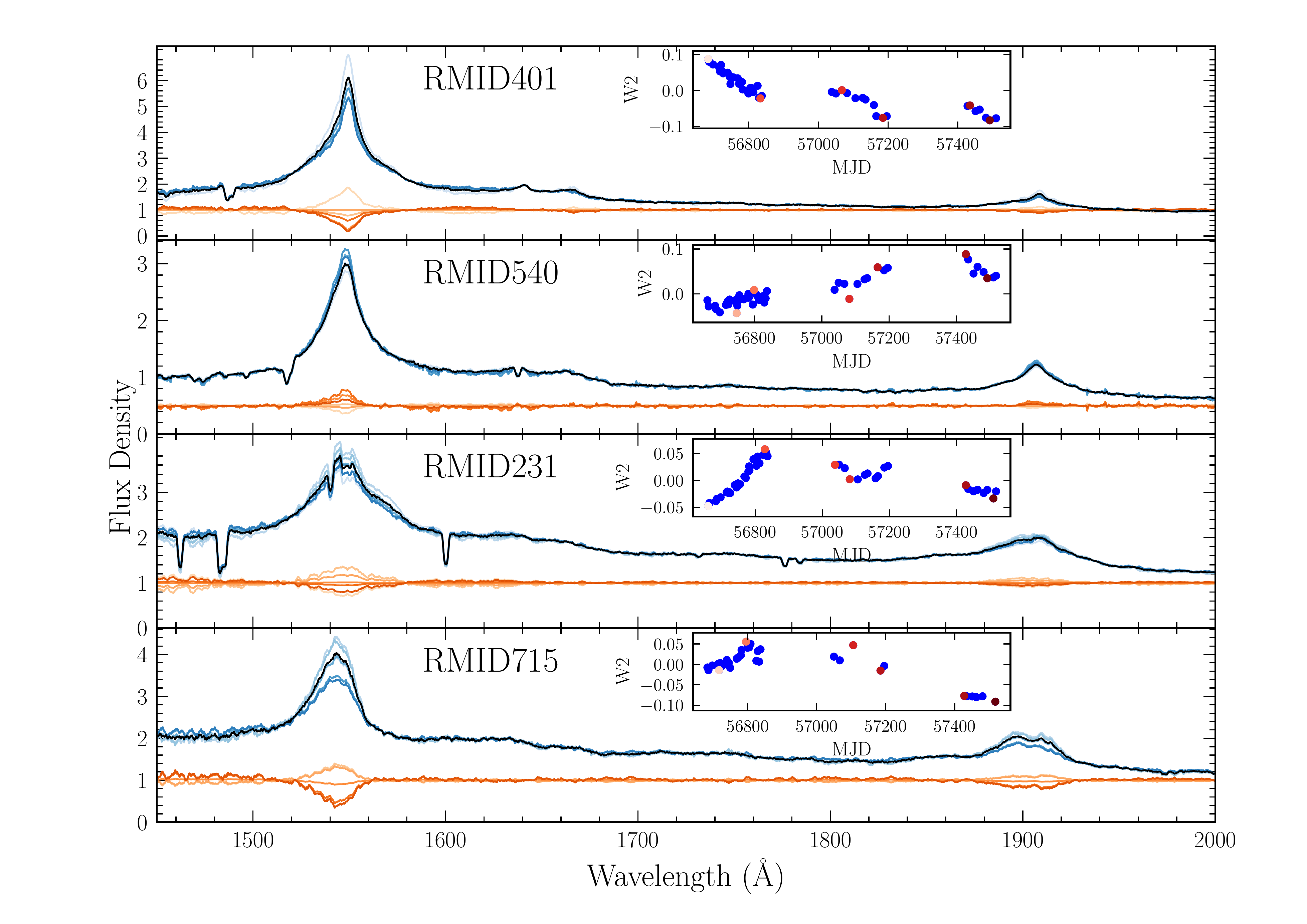}
    \caption{The median ICA reconstruction (black) with reconstructions spanning a range of W2 values overlaid (shades of blue, lighter colors indicating earlier epochs) for four objects representing the different areas of \civ\ parameter space; see \S~\ref{sec:disc}. In order: RMID 401 (representing large \civ\ EQW and low blueshift), RMID540 (intermediate EQW and blueshift), RMID231 (low EQW and blueshift), and RMID715 (low EQW and high blueshift). The various W2*Component 2 arrays are shown in red hues.  The inset shows how W2 changes with time; the spectral epochs plotted in the main panel are indicated with the red-hued points (light to dark over time).
    \label{fig:fig4}}
\end{figure*}

Figure~\ref{fig:fig3} shows two example objects with coverage of the Ly$\alpha$, \ion{Si}{4}, \civ, \ciii, and \mgii\ emission lines, plotting both Components 1 and 2 (with the latter shown on an enlarged scale indicated on the right axis) at two different epochs (one where W2 is most negative and one where it is most positive).  We find that all of the emission lines appear together in Component 2 and that the Component 2 emission-line region is roughly similar in shape to the emission-line regions of Component 1.  The first observation means either that the emission lines are varying together and/or that the continuum is dominating the changes in line strength (see Section~\ref{sec:eqw} for further discussion).  The second observation means that there are no strong velocity-dependent changes in the line shape with time.
Figure~\ref{fig:fig4} zooms in to the \ion{C}{4} through \ion{C}{3}] region for four example objects that span the range of emission-line properties seen in the sample.  As insets we illustrate the weight of Component 2 (W2) with time, highlighting six epochs in time with different shading. 
RMID401 has an observed-frame time lag of 134--171 days \citep{Grier+2019} such that there is no direct information about the behaviour of the continuum and the BELR at the same time.  RMID231 and RMID540 are also shown in Section \ref{sec:CombRecon} and are examples of the more variable objects in the sample.

\begin{figure}[th!]
\includegraphics[width=3.5in]{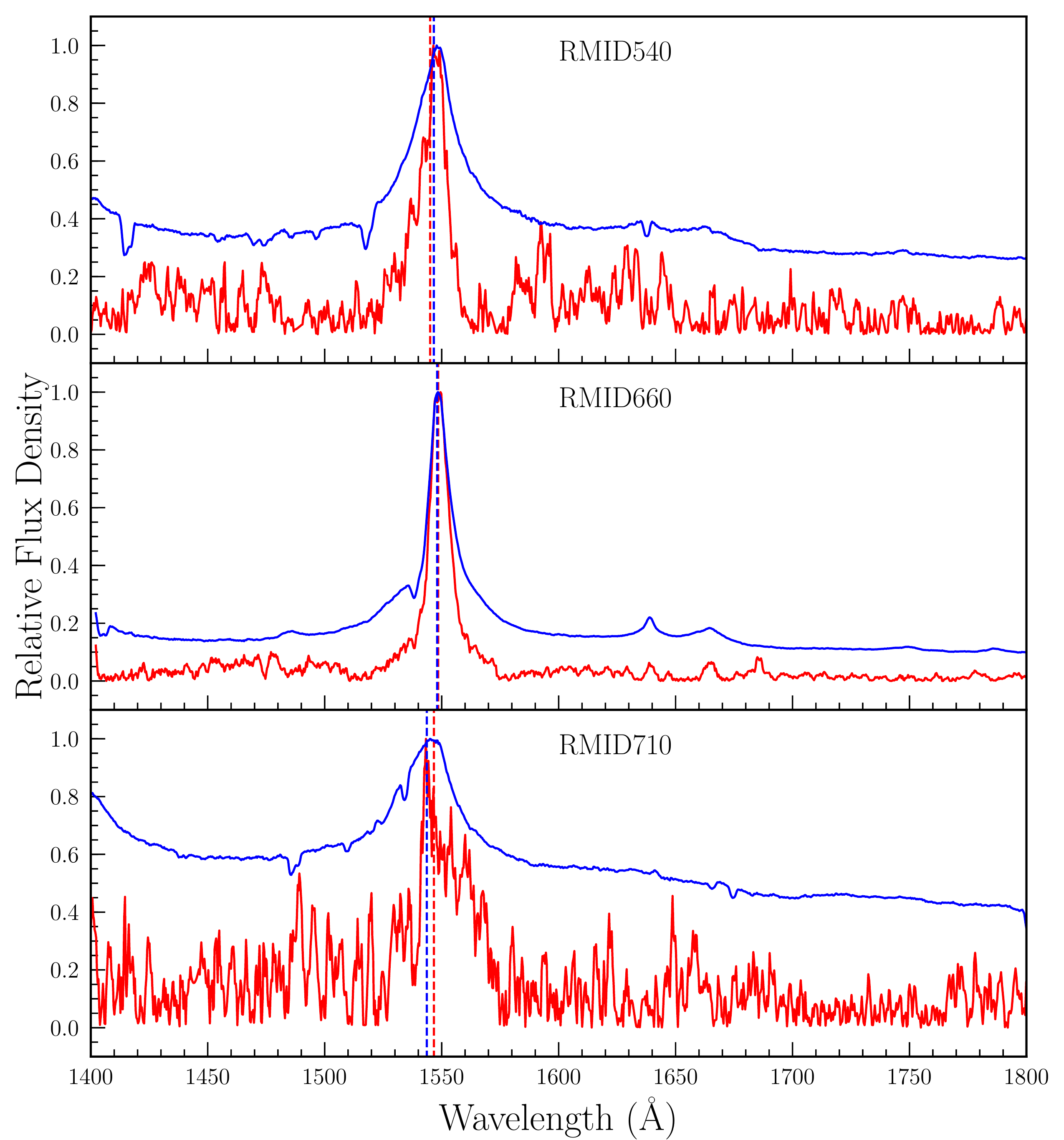}
\caption{Component 1 (blue solid line) in comparison to Component 2 for that same observation (red solid line). The half-flux wavelengths for both are illustrated by the vertical dashed lines (blue and red, respectively). In the top panel, the flux of RMID 540's Component 2 is shown to be skewed blueward relative to Component 1. In the middle panel, RMID 660's Component 1 and Component 2 possess their half-flux wavelengths at close to the same wavelength. In the bottom panel, RMID 710's Component 2 is skewed redward relative to Component 1.
\label{fig:fig5}}
\end{figure}

Figure~\ref{fig:fig5} illustrates three possibilities in terms of the shapes of Component 2 relative to Component 1.  In the top panel, we show an example where Component 2 has an asymmetry blueward of Component 1.  In the middle panel, we show Component 2 having a similar shape as Component 1.  In the bottom panel, we show an example of Component 2 that is redward asymmetric as compared to Component 1.  The second case is what we would expect if the changes in the line strength are dominated by changes in the continuum (but this can also happen if the BELR is really changing).  The first and third cases, on the other hand, represent objects where the BELR {\em must} actually be changing.  If the \civ\ BELR is made up of multiple components, and they are reacting to the continuum with different lags times (due to different characteristic differences), we might expect Component 2 to have a different shape than Component 1. As we define the component weights to be positive when the EQW is increasing with time, objects like those in the top panel of Figure~\ref{fig:fig5} with blue asymmetric Component 2s by definition will have a positive correlation between blueshift and EQW, while those objects with red asymmetric Component 2s will have a negative correlation between blueshift and EQW.  Thus all of the information about how these parameters relate to each other with time for an individual object are almost completely encoded in the relative shapes of Components 1 and 2.  See Section~\ref{sec:CIVCorrelations} for further discussion.

When the time sampling is long enough and the cadence is frequent enough we would expect to see correlated changes between the lines and also velocity dependent changes across individual lines (both due to time lag effects). An example of the latter is if the flux changes in the wings are different than the changes in the line core. 
The presence of such changes are conducive to determining the lag times needed for reverberation-mapping analysis \citep{Grier+2019}. 

\begin{figure}{}
    %\epsscale{1.2}
    \includegraphics[width=3.5in]{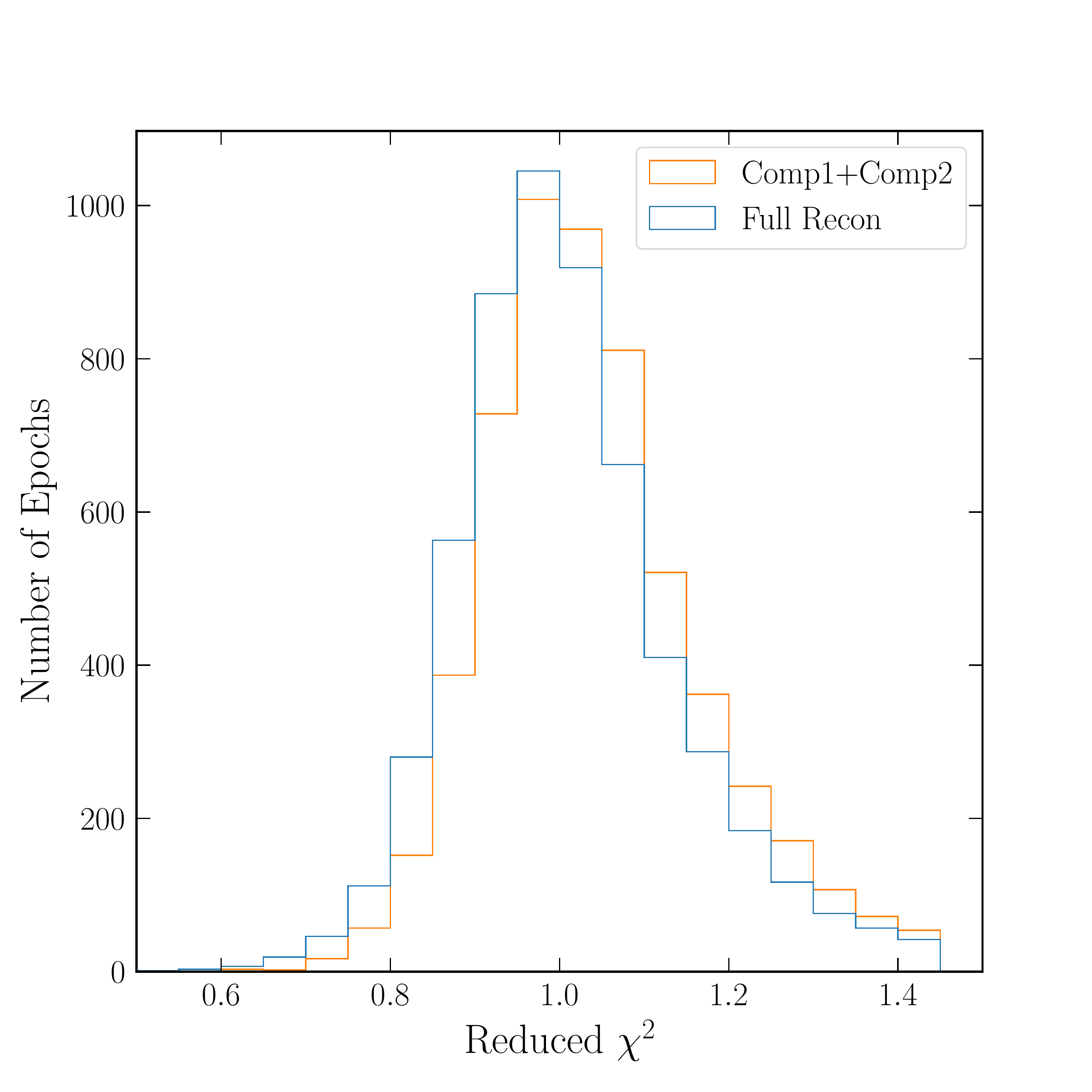}
    \caption{Histograms of the reduced $\chi^{2}$ values calculated from comparing our processed DR14 data with the individual ICA reconstructions including 2 (orange) or 3 (blue) components. There is some improvement when using the full reconstruction, however in most cases two components is adequate.
    \label{fig:fig6}}
\end{figure}

In such cases the ICA analysis requires a third component in order to reproduce changing emission-line profiles with time. For the majority of the quasars a third ICA-component was found to unnecessary, as determined from a reduced $\chi^{2}$ analysis, see Figure \ref{fig:fig6}. The need for just a single varying component is a significant result in and of itself.  For example, it says that the line changes over the wavelength range covered (Ly$\alpha$ to \ion{Mg}{2}) are highly correlated.  Indeed, this property is exactly what makes the ICA reconstruction technique so powerful.  Since the other UV emission lines are correlated with \ion{C}{4} emission, the ICA algorithm is able to reconstruct the intrinsic spectrum even when there are strong absorption features.  Later we will use this property to improve the S/N of line parameters extracted from single-epoch spectra (\S~\ref{sec:linecalc}).

\subsection{ICA Reconstructions as f(t)}
\label{sec:CombRecon}

\begin{figure}
\includegraphics[width=3.5in]{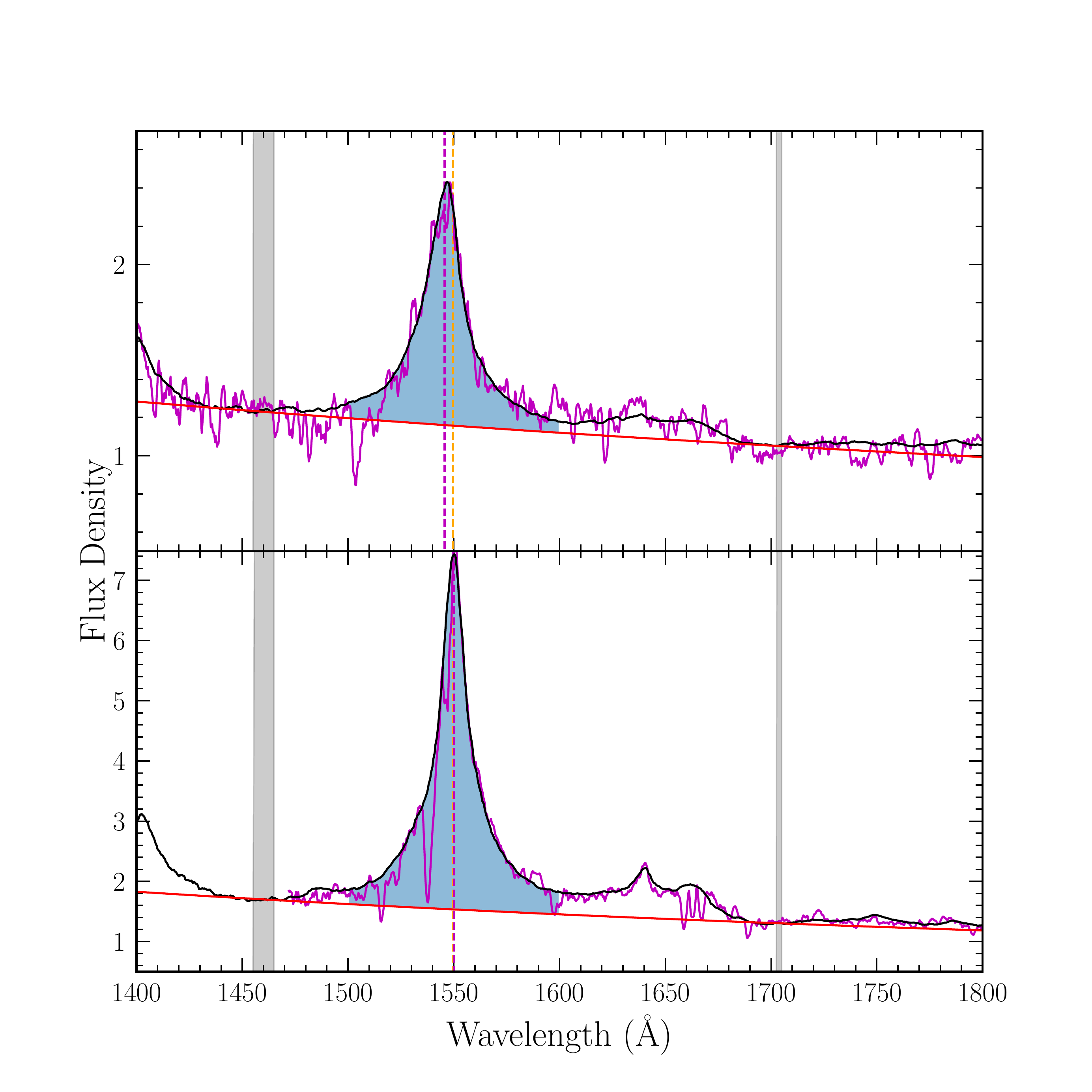}
\caption{Illustration of the quality of the combined global+individual ICA reconstructions.  Top: RMID178, Epoch 44. Bottom: RMID321, Epoch 1. The original data is represented by the purple solid line, the ICA reconstruction by the black solid line. The continuum is shown in red, with the windows used to fit that continuum shown in gray. The laboratory and half-flux wavelengths are also shown, by the dashed orange and purple lines, respectively. 
\label{fig:fig7}}
\end{figure}

``Absorber-free" ICA reconstructions for each epoch of each quasar were obtained via a two-stage procedure. First, the global reconstruction of the median reconstruction of a quasar---which represents an absorber-free version of the spectrum (Section \ref{sec:globalica})---is ``unmorphed" to preserve the large-scale shape of each epoch (Section \ref{sec:dataproc2}). Components 2 and 3 from the individual ICA (Section \ref{sec:individualica}) are then weighted (by W2 and W3) and added to the unmorphed global reconstructions. An example of this type of reconstruction and the corresponding original data is shown in \ref{fig:fig7} by the purple and black lines, respectively.

\section{Extraction of Line Parameters}
\label{sec:lineparams}

\subsection{Equivalent Width versus Line Flux}
\label{sec:eqw}

Before extracting line parameters we discuss how such parameters are measured.  It is first important to recall the definition (and potential shortcomings) of both EQW \citep{Peterson1997} and line flux as ways to parameterize the strength of an emission line. We start by comparing the definition of line flux to that of EQW.

To quantify the flux produced by line-emitting gas in an emission-line object we compute the integrated line flux as

\begin{equation}
\int^{\lambda_{\rm max}}_{\lambda_{\rm min}} (F_{\rm tot} - F_{\rm cont}) \delta \lambda \simeq \Sigma (F_{\rm tot} - F_{\rm cont}) \Delta \lambda.
\end{equation}
For example if we have a continuum level of 10 (in arbitrary units) that is uniform in wavelength and a top-hat emission line with height of 20 that spans 10 bins of 1 \AA\ each, then the total line flux would be (20-10)(10)(1) = 100.

If, however, the spectrophotometry is not well determined (e.g., due to dome obscuration, \citealt{Kent1985}, or an optical fiber falling out in mid exposure, \citealt[\S~3.3]{SDSSRM}), then we must instead compute the equivalent width, normalizing by the continuum:

\begin{equation}
{\rm EQW} = \int^{\lambda_{\rm max}}_{\lambda_{\rm min}} \left( \frac{(F_{\rm tot} - F_{\rm cont})}{F_{\rm cont}}\right) \Delta \lambda.
\label{eq:eqw}
\end{equation}

In our example above, the EQW would then be ((20-10)/10)(10) = 10\AA.  If our exposure only included 50\% of the light for some technical reason, then the measured line flux would be (10-5)(10) = 50, while the EQW would be ((10-5)/5)(10) = 10\AA\ as it was before.  Thus the EQW is a more appropriate measure of the total line flux in the case of uncertain spectrophotometry---as is the case for SDSS-III spectra \citep{Margala+16}.\footnote[4]{Although SDSS-RM used more sky fibers and drilled for a different wavelength, so this should be less problematic.}

In the case where the continuum emission is coming from a very different region than the emission-line flux \citep{Pogge+1992}, then the continuum that we should reference is not the continuum observed at the same time as the emission-line flux, but rather the continuum at an earlier time, specifically earlier by the light-crossing time between the (characteristic) radii of the continuum and emission-line regions.

In the example presented by \citet{Pogge+1992}, investigating the ``intrinsic" Baldwin Effect (how EQW changes with $L$ for a given object, rather than for the population), the lag is only 8 days---much shorter than the time over which the light curve was measured. Thus this lag merely introduces a small error in the EQW.  
As discussed in Section~\ref{sec:sample}, in the case of the SDSS-RM objects the lags are much longer and normalizing by the continuum observed at the same time as the emission lines does not simply introduce a small error, rather the continuum potentially dominates changes in EQW.  That is, if the continuum is changing rapidly, the EQW will be computed as also changing rapidly---even if the line flux did not change at all during the same time-lagged period. In such a case, we expect the shapes of Components 1 and 2 to be the same (see Figure~\ref{fig:fig5}).  Thus, without accurate determination of lags (and observations of both the continuum and BELR at the appropriately lagged times), it is not possible to use quasars with long lags to investigate the intrinsic Baldwin Effect.  

Moreover, it is generally not the continuum flux at the line center that we want to know, but rather the continuum at the energy of creation for the species in consideration.  For \ion{C}{4} we do not want the 1550\AA\ continuum, but rather the continuum at 48eV.  We cannot reference the 48eV continuum (rather line changes can be used to infer changes in the 48eV continuum), so we adopt standard procedure and assume that the 1550\AA\ continuum is sufficient.

The fact that the measurement of EQW is dependent on the time-variable continuum suggests that line flux may be a better indicator of the actual changes in the line strength.  However, line flux is problematic in a different way.  Specifically, unlike EQW, it is not normalized for the luminosity.  That means line flux is a good parameter for comparing the line strength of one quasar over the course of time, but it is not a good measurement to use to compare different quasars: more luminous quasars will have inherently larger line strengths.

As a result, in addition to EQW and line flux, we will adopt a third measurement of the line strength, which we will refer to as the luminosity normalized EQW (or normedEQW).  For the normedEQW instead of using a wavelength- and time-dependent continuum in the denominator of Equation~\ref{eq:eqw}, we instead use the median monochromatic continuum luminosity measured at 1550\AA\ (derived from the Custom+PrepSpec reductions).  The normedEQW will therefore not be sensitive to continuum variability that is not (currently) affecting the line strength and can be used to provide a robust measure of the time-dependent line strength in a single quasar.  The normedEQW will also be normalized by the luminosity of that quasar, meaning that it can be used to compare quasars of different luminosity. Which of these three line strength measurements (line flux, EQW, normedEQW) are used will depend on the nature of the analysis.

\subsection{Single-epoch versus Multi-epoch Data}

While we cannot actually learn anything about changing \civ\ flux from objects where the EQW changes are dominated by {\em continuum} changes (to which the line-emitting gas has not had time to respond), the changes seen nevertheless reflect what we would measure at any given snapshot in time.  That is, single-epoch spectroscopy (or multi-epoch where the time lag is not known) {\em inherently} assumes that the emission line flux and continuum flux are contemporary.  
As a result, this data set is ideal for determining the effect of non-contemporaneous continuum variability on the measured EQW.  As our analysis was performed on the standard BOSS spectral reductions that were applied to the SDSS-RM spectra, our process can be applied to any of the single-epoch BOSS spectra equally and not just those observed as part of the RM program.  

On the other hand, we will use the Custom+PrepSpec calibrations (\S~\ref{sec:dataproc2}) when determining the median luminosity of the objects in our sample and for determining the line fluxes and normedEQWs.  The difference between these and the BOSS calibrations are not important for the EQWs (essentially by definition of the EQW) or for the blueshifts.   Thus those parameters will be measured from the BOSS reductions to allow comparison with the larger BOSS sample of quasars. 

\subsection{Calculating Line Parameters}
\label{sec:linecalc}

Whether we are computing line flux, EQW, or normedEQW, we adopt the same procedure for extracting these properties from the spectra.  First, we make measurements from the ICA reconstructions from Section~\ref{sec:CombRecon} rather than from the observed spectra.  Doing so is justified by the results of Figures~\ref{fig:fig6} and \ref{fig:fig7} showing that the ICA reconstructions are an accurate representation of the spectra. We similarly measured the EQW and blueshift directly from the original spectra to enable comparison. We follow the methods of \citet{Coatman+2017} in using a local power-law continuum fit to the \civ\ emission-line region using the default continuum windows of [1445, 1465] \AA\ and [1700, 1705] \AA, but we adjusted these when necessary (e.g. to avoid absorption features or to allow a wider window for lower S/N). For the calculations of EQW and blueshift we used an integration window of [1500, 1600] \AA,  except in the case of high blueshift objects, in which the lower edge of the window was changed to 1460 \AA\ and the upper edge of the window was identified by eye. To determine the blueshifts we use:
\begin{equation}
\textrm{\civ\ Blueshift (km $\textrm{s}^{-1}$)} = c*(1549.48-\lambda_{half})/1549.48
\end{equation}
where 1549.48 $\textrm{\AA}$ is the rest-frame wavelength for the \civ\ line and $\lambda_{half}$ is the wavelength bisecting the cumulative flux.  This process for deriving the \civ\ EQW and blueshift was the same as was used in \citet{Rankine+2020}.  The continuum and integration windows along with the half-flux wavelength were illustrated in Figure~\ref{fig:fig7}. Errors were calculated by using a Monte Carlo approach. The continuum fit for each epoch of each object was perturbed by some random contribution of the noise array 50 times, then EQW and blueshift measurements were made using these perturbed continua. The standard deviation of the resulting EQWs and blueshifts was taken to be the 1-$\sigma$ error on the measurements for that object. It is important to note that there are additional sources of error which are not accounted for by this process, such as how absorption features still have an affect on components 2 and 3, and how well the ICA reconstruction recreated the original spectra. 

\begin{figure}
\includegraphics[width=3.5in]{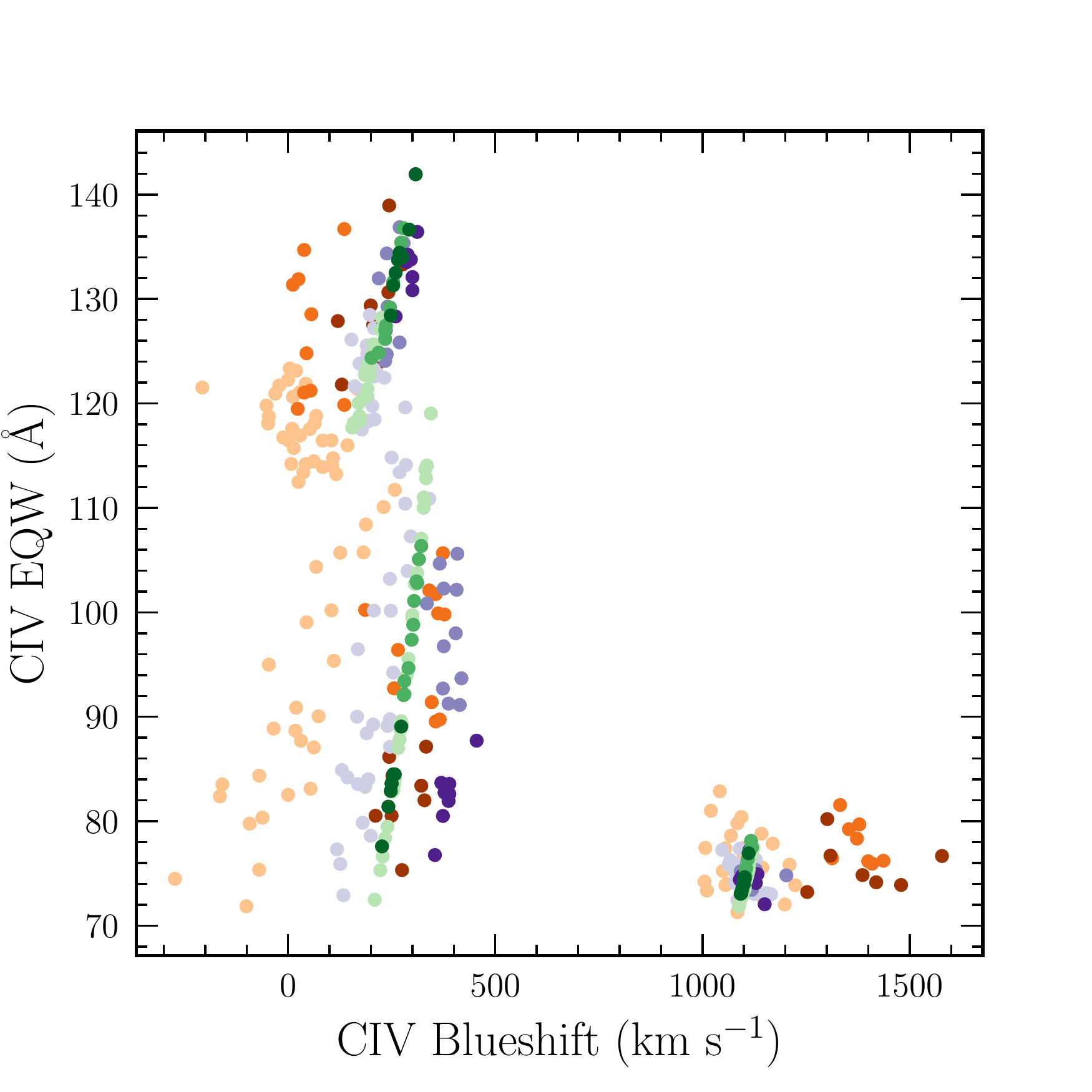}
\caption{The \civ\ parameter measurements for RMIDs 231, 540, and 713 (bottom left, top left, and bottom right, respectively). The points are colored by year, earlier seasons are lighter colors. The orange points represent measurements from the original spectra, the green points measurements from the absorption-corrected ICA reconstructions with W2*Component 2 from the individual ICA. The purple points are measurements from the absorption-corrected ICA reconstructions with both W2*Component 2 and W3*Component 3 added. Examples of Component 2 and the ICA reconstructions for \civ\ and CIII] for RMID231 and 540 were previously shown in Figure~\ref{fig:fig4}.
\label{fig:fig8}
}
\end{figure}

\begin{figure}
\includegraphics[width=3.5in]{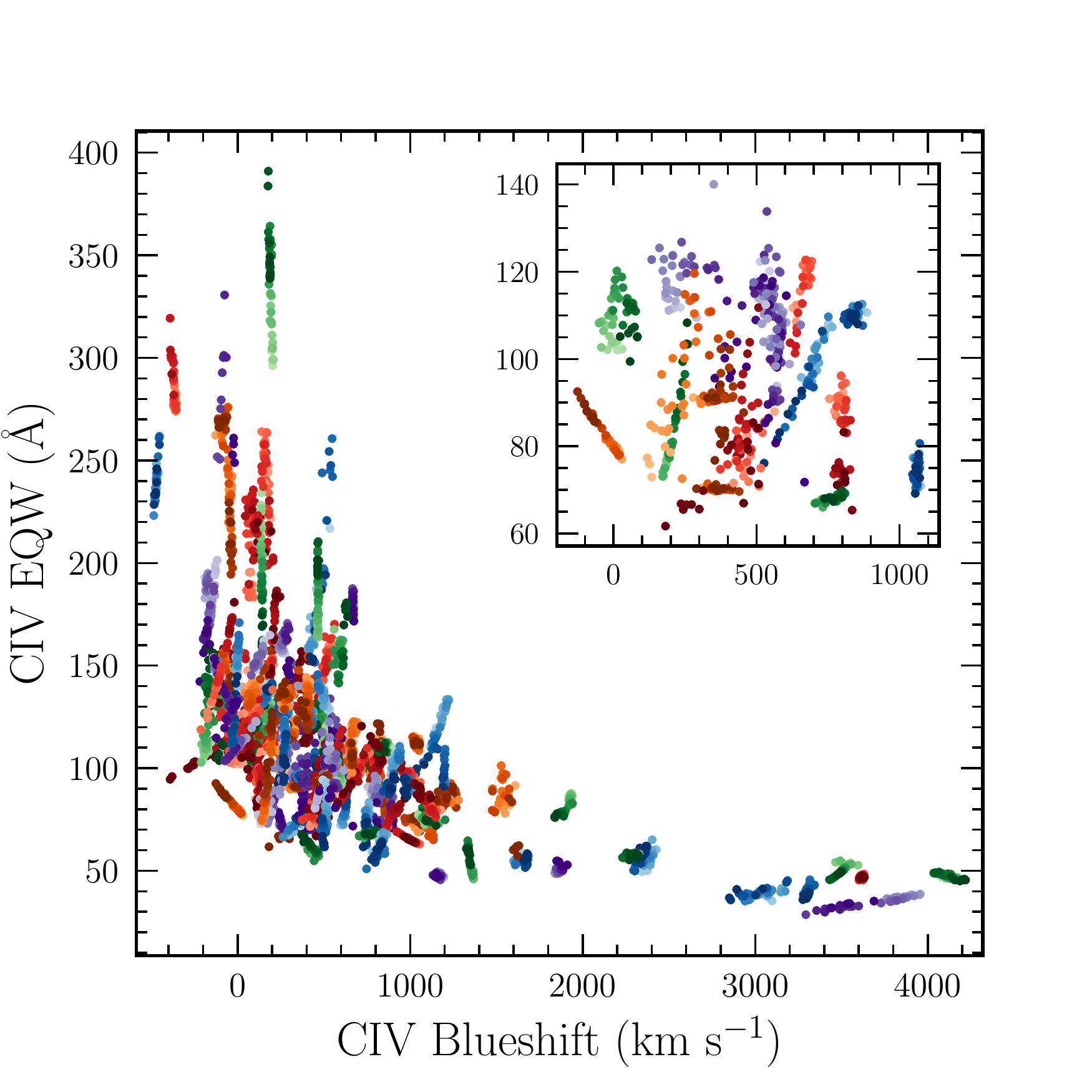}
\caption{The \civ\ parameters as calculated with the global ICA reconstruction with individual Components 2 and 3 added for all of the quasars in our sample.  Colors are not meant to group objects, but different shades indicate measurements from different epochs (lighter colors are earlier in time). A sample of objects from the lower left hand corner of the distribution is magnified in the inset.
\label{fig:fig9}
}
\end{figure}

What is novel herein is that the ICA reconstructions make use of all of the line information. In Figure~\ref{fig:fig8} we plot EQW versus blueshift for RMIDs~231, 540, and 715.  RMID~231 is the object in the sample that has the largest contribution from Component 3.  Orange hues indicate the EQW and blueshift values derived from the original spectra for each epoch, while purple hues indicate the values derived from the full ICA reconstructions, showing the improvement in S/N inherent to the method.  Overlaid in green hues are the parameters derived from a reconstruction only including Components 1 and 2.  Unlike these three objects, most quasars in the sample have negligible contribution from Component 3 and have less variable contributions from Component 2. These objects were chosen from a subsample of ``clean" objects where there is nothing about the spectral features (e.g., absorption) or the data reduction (BOSS versus custom) that are expected to change the results. This was done so that the differences between parameters measured from the observed and reconstructed spectra are dominated by the improvement of our process and not by those features. We found that the \civ\ parameters are much more localized when determined from the ICA reconstructions (effectively increasing the S/N of the spectra). The detailed reason for this improved S/N is that there are strong correlations between \civ\ and the other UV emission lines, particularly \ion{He}{2} and the \ion{C}{3}] line complex (which are observable for all of the reshifts investigated herein); see   \citet[][Fig.~16]{Richards+2011}, \citet{Baskin+2013}, and \citet{Rankine+2020}.  Thus, the accuracy of the extracted line parameters will be more robust than is otherwise indicated by the spectral S/N at each epoch. 
Figure~\ref{fig:fig9} shows the distribution of EQW versus  blueshift for {\em all} the quasars in our sample.  Here we plot the \civ\ measurements derived from the full ICA reconstructions and for each epoch (i.e., equivalent to the purple-shaded points in Figure~\ref{fig:fig8}). 

Had we been able to perform ICA reconstruction with only a single set of components for all objects, then there would be no need to measure parameters of individual lines at all.  Instead we could investigate changes in the {\em bulk} emission line properties using only the ICA weights.  While that remains a goal for future work, in this paper we will concentrate on what can be learned from traditional line parameters as discussed in the previous section. 

We note that, for a Gaussian, the EQW is simply related to the product of the height and the FWHM. Thus there are two parameters that can change the EQW. Indeed in some cases (RMID275 being a prime example), we find that changes in the FWHM significantly affect changes in the EQW (see also \citealt{Wilhite+2006}).  That finding is interesting given that the FWHM is the main parameter used to estimate BH masses from scaling relations and the BH masses cannot be changing significantly over the timescales sampled.  Indeed, \citet{Wilhite+07} and \citet{Park+12} find that $\sim 12$--15\% of the error in BH mass estimates is due to variability.  See work on BELR ``breathing" by \citet{Wang+2020} for further discussion of line width changes in the SDSS-RM data set.

\section{Analysis/Discussion}
\label{sec:disc}
    
\subsection{Variability}
\label{sec:varintro}

We would like to be able to describe each quasar by its properties along some sort of ``main sequence" \citep[e.g.,][]{Marziani+2018}.  As discussed above, at low redshift the Eigenvector 1 correlates provide such a parameter space, while at high redshift, the \civ\ line can be utilized.  However, regardless of the redshift of a quasar, measurement error and variability fundamentally distort this mapping.  That is, measurements that are needed to determine fundamental properties like BH mass and accretion rate will give different results in different epochs, even if the fundamental properties have not actually changed.

The SDSS-RM sample enables exploration of the reliability of single-epoch measurements. We will explore four facets of the problem: (1) how localized quasars are in the \civ\ parameter space, (2) the degree to which changes in EQW and blueshift are correlated (3) how intrinsic variability changes across the \civ\ parameter space, and (4) how that variability affects determination of the Baldwin Effect slope from single-epoch spectra.

\subsubsection{Localization in \civ\ Space}
\begin{figure}
\includegraphics[width=3.5in]{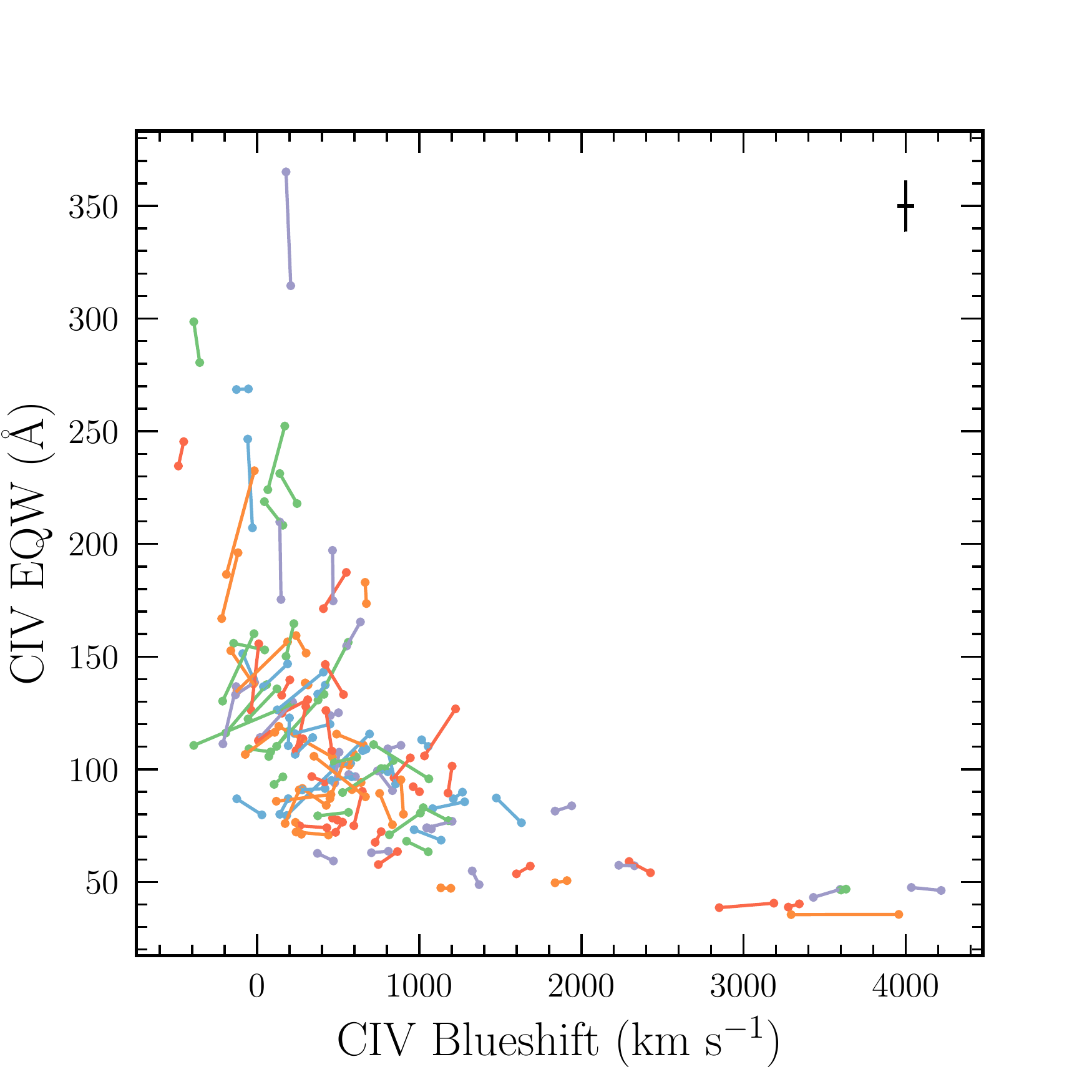}
\caption{The \civ\ normedEQW and blueshift parameter pairs showing the greatest change between different epochs. The black cross in the upper right hand corner shows the median 2-$\sigma$ errors for the \civ\ EQW and blueshift for the sample. Note that there are no objects crossing from one side of the parameter space to the other.
\label{fig:fig10}
}
\end{figure}

Figures \ref{fig:fig9} and \ref{fig:fig10} provide a global perspective for understanding the degree of localization within the \civ\ parameter space. Figure~\ref{fig:fig10} reproduces Figures~\ref{fig:fig9}, except it shows only the two most extreme epochs and was created with normedEQWs.  We find that there are no examples of objects crossing from one side of the parameter space to the other (on rest-frame timescales of $\sim$ 300 days), as was also noted in \citet{Sun+2018} using a similar sample.  This observation says that, at least to first order, single-epoch measurements of EQW and blueshift give a good indication of the relative location of an object in the \civ\ parameter space.  

To the extent that the location depends on \lledd,
then single-epoch \civ\ parameters alone may be good proxy for \lledd.  If that is the case, and if the X-ray to optical spectral index and in-band X-ray spectral index are also good indicators of \lledd\ \citep{Shemmer+2006,Marlar+2018}, then we might expect the \civ\ and X-ray properties of quasars to be correlated \citep{Gallagher+2005, Kruczek+2011, Ni+2018,Rivera+2020} and for the combination of \civ\ and X-ray parameters to yield more accurate \lledd\ than either method alone.

\subsubsection{Correlations in \civ\ Space}
\label{sec:CIVCorrelations}

\begin{figure}
\includegraphics[width=3.5in]{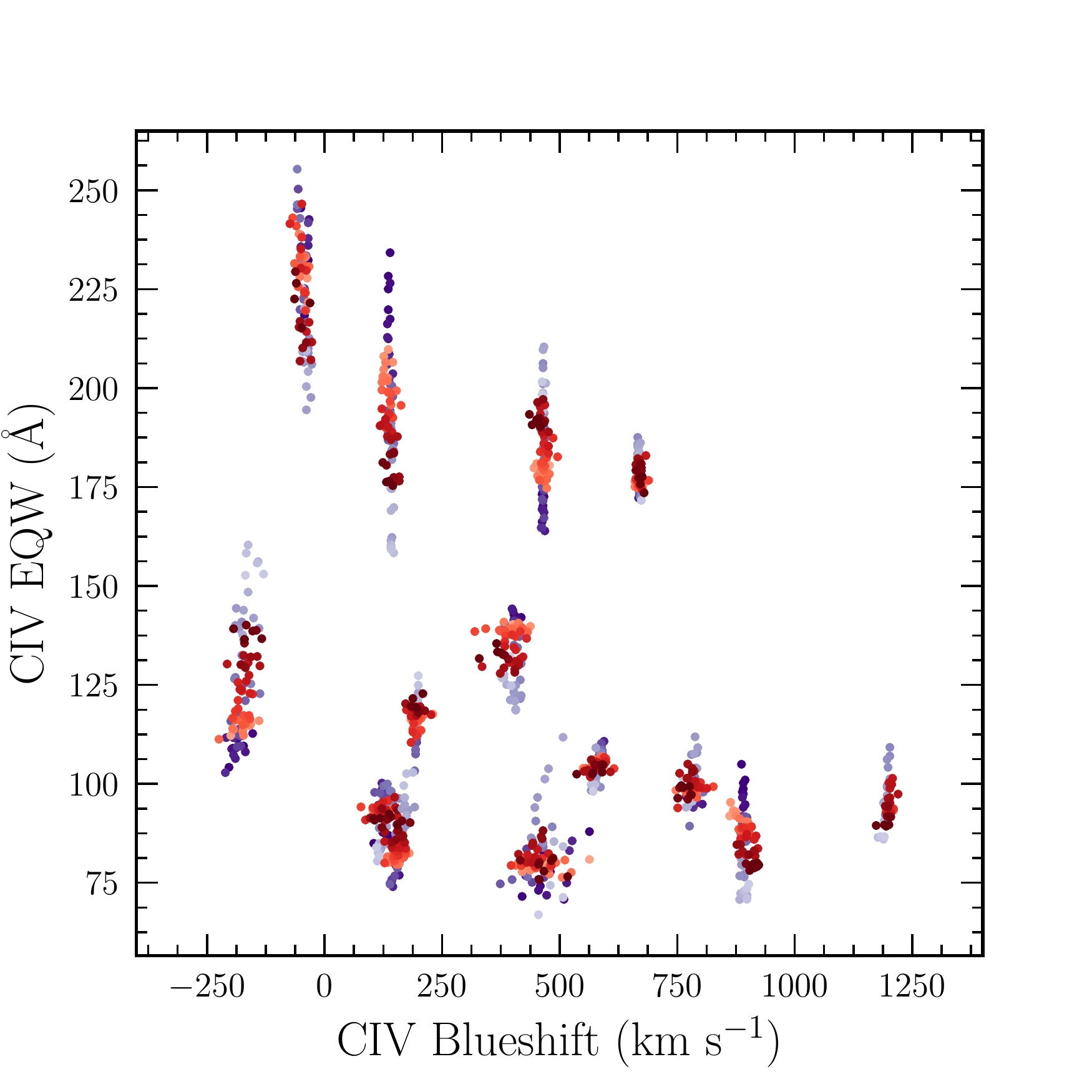}
\caption{The \civ\ parameters for vertical objects in the clean sample. Purple points are ICA EQWs, red are the normedEQWs. The purple points are now reversed in time to prevent dark colors from overlapping. Note these objects are not as variable in normedEQW, however there is still variability in the strength of the emission line in those objects.
\label{fig:fig11}
}
\end{figure}

We next explore the degree to which changes in EQW and blueshift are correlated (or uncorrelated).
\citet{Sun+2018} argued that there is an overall negative correlation (positive using their sign convention) such that epochs with smaller blueshifts have larger EQWs, which enhances the global anti-correlation.  
In exploring these relationships, two aspects of the data must be kept in mind: the effect of unlagged continua (\S~\ref{sec:eqw}) and the shape of Component 2 relative to Component 1 (\S~\ref{sec:individualica}).  Comparing EQW to normedEQW as shown in 
Figure~\ref{fig:fig11} illustrates how much not having rest-frame contemporaneous measurements of the BELR and continuum matters. In this figure we limited the analysis to ``clean" objects. The red hued points are the normedEQWs, while the purple points are the EQWs, but reversed in time order to prevent darker colors from overlapping.  This presentation suggests that up to half of the EQW ``variability" can be caused simply by changes in the continuum.  Furthermore differences in the shapes of Components 1 and 2 as illustrated in Figure~\ref{fig:fig5} (e.g., peaking at different wavelengths or having different skewnesses) mean that there will be changes in blueshift with time.  It is then possible for there to be {\em either} a positive or negative correlation between blueshift and EQW as a result of unlagged continuum changes---even if the EQW is not actually changing at all, since the EQW can be responding to the current continuum while the line itself (as characterized by the blueshift) is responding to the lagged continuum.  With those caveats in mind, we consider four relationships and speculate about what physical meaning they might have in the context of the wind model of \citet{Giustini+2019}:
\begin{itemize}
\item Vertical (changing EQW, but constant blueshift) may first indicate significant continuum variability as discussed above and in Section~\ref{sec:eqw}.  However, a residual vertical trend after normalizing the EQW (Figure~\ref{fig:fig11}) would suggest that there is little wind contribution (such as is the case for low BH mass, moderate accretion rate quasars in Fig.~1 of \citealt{Giustini+2019}) and that the ionization state is simply changing the volume of \civ\ emitting gas (perhaps in the form of a failed wind).
\item Horizontal (changing blueshift, but constant EQW) may be a sign that wind is dominating the BELR, with a relatively low volume of \civ\ emitting gas in a failed state, consistent with the very high accretion rate sources in Fig.~1 of \citet{Giustini+2019}.  In this case it may be only the gas located at the transition between a wind and failed wind that is affected---without any actual changes to the volume of \civ-emitting gas.
\item A residual negative correlation (using normedEQW or properly lagged EQWs, where the relative shapes of Components 1 and 2 must be like the bottom panel of Figure~\ref{fig:fig5}) might be expected for the majority of sources given that such a correlation tracks with the general trend (of lower EQW with larger blueshift).  A negative correlation might be expected in the picture of \citet{Giustini+2019} in the case of increasing/decreasing accretion rate.  For example, moving from moderate to high accretion rate in their Fig.~1 (at either low or high mass), would both decrease the volume of \civ-emitting gas in the failed wind component (thereby reducing the EQW), but also decrease the radius where the wind is launched (thereby increasing the maximum terminal velocity of the wind). 
Such changes in effective accretion rate need not be significant (only systematic) in order to produce such an effect.
\item A residual positive correlation (top panel of Figure~\ref{fig:fig5}) is somewhat harder to understand in this context as it would say that both the volume of gas in the failed wind is increasing and that the minimum launch radius is decreasing.  It would therefore be interesting to look for other explanations for such systems.  For example, if there is evidence in these objects that the \civ\ EQW is contaminated by flux from another line (e.g., iron emission, \citealt{VW01}), as might be the case where there is a significant ``shoulder" between \civ\ and \ion{He}{2}.
\end{itemize}

\subsubsection{Variability Across \civ\ Space}
\label{sec:var}

\begin{figure}
\includegraphics[width=3.5in]{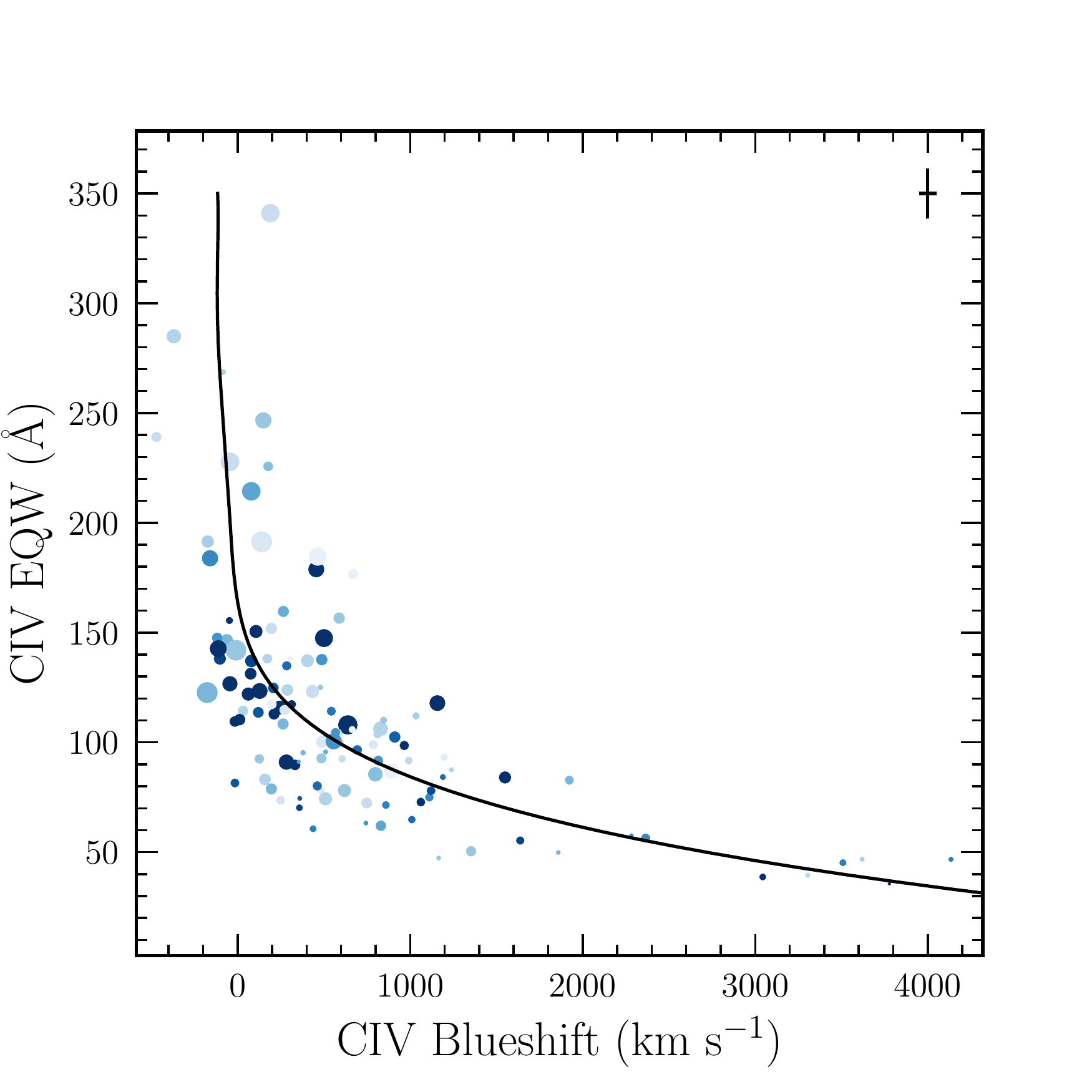}
\caption{Median blueshift and normedEQW for the full sample. The black cross in the upper right hand corner shows the median 2-$\sigma$ errors for the \civ\ EQW and blueshift for the sample. The points are color-coded by the MAD of the \civ\ blueshift with sizes that indicate the MAD of the normedEQW.  Small, light-colored points are not very variable over the timescales sampled.  High-blueshift quasars are not very variable in EQW, while high-EQW quasars are not very variable in blueshift. 
\label{fig:fig12}
}
\end{figure}

We can characterize how much the objects vary in a number of ways.  In Figure~\ref{fig:fig12} we simplify Figure~\ref{fig:fig9} by plotting only the median values of normedEQW and blueshift for each object.  Using these data points we construct a best-fit curve using polynomial fitting to help describe the data trends.  For each object, we can compute the closest distance to this curve and then describe that object according to the distance ``along" the curve, 
thus creating a sort of first principal component.  Identification of trends may be facilitated by considering the sample using this ordering; as such, the figures herein are sorted according to this distance where appropriate.

\begin{figure}[h!]
\includegraphics[width=3.5in]{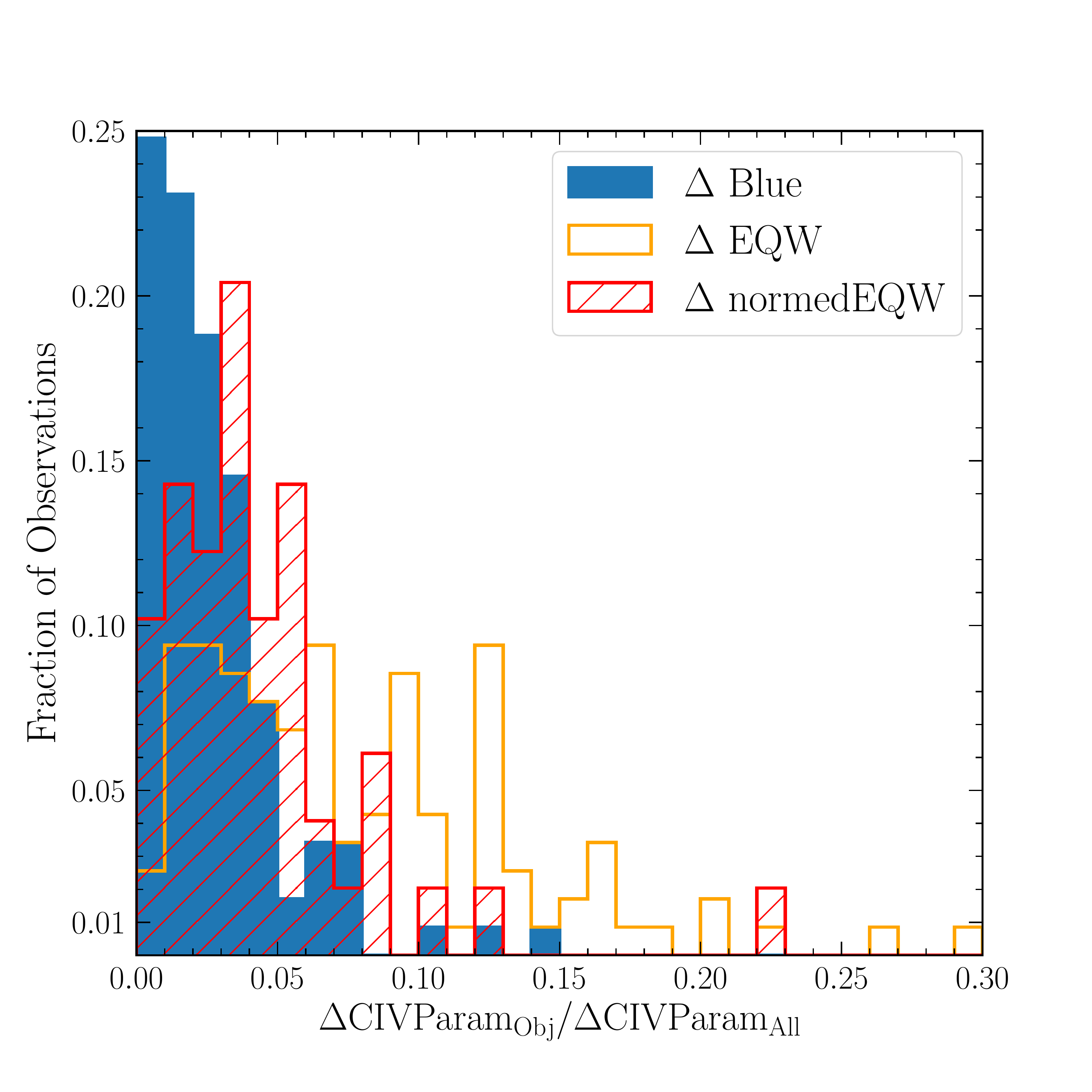}
\caption{The maximum change in an object's EQW, normedEQW, or blueshift over the maximum change in the sample's blueshift, EQW, or normedEQW, (solid blue, open orange, and hatched red, respectively). We see that the fractional change in blueshift is generally much less than the fractional change in EQW.}
\label{fig:fig13}
\end{figure}

We express the variance of normedEQW and blueshift in Figure~\ref{fig:fig12} by changing the plot marker to reflect the median absolute deviations (MAD) of the blueshift (darker hues being more variable) and normedEQW (larger points being more variable).  
Figure~\ref{fig:fig13} illustrates the same result in a different way.  Here we compute the ratio of the range of each individual object compared to the range of all objects.  For example, if the EQW of an object changed by 60\AA, and the full sample spans a range of 300\AA, then the ratio is 0.2.  We compute this ratio for both EQW and blueshift as it results in a unitless metric which is on the same scale for both parameters (unlike the MAD).  

From both of these plots, we see that the blueshift is much less variable than the EQW, which is expected if the variance of the EQW is affected by the continuum.  We can recompute this metric for the clean objects as noted above, using normedEQW, which should eliminate the artificial variability of EQW.   This result is shown by the red histogram in Figure~\ref{fig:fig13}. The more concentrated result confirms that the continuum is affecting the EQW measurements, but also suggests that (true) EQW is indeed more variable than blueshift.
In the context of a two-component model of the BELR, this finding may simply suggest that the volume of \civ-emitting gas in the disk component generally dominates the wind component, such that it is typically difficult to observe changes to the wind component.

\begin{figure}{}
    \includegraphics[width=3.5in]{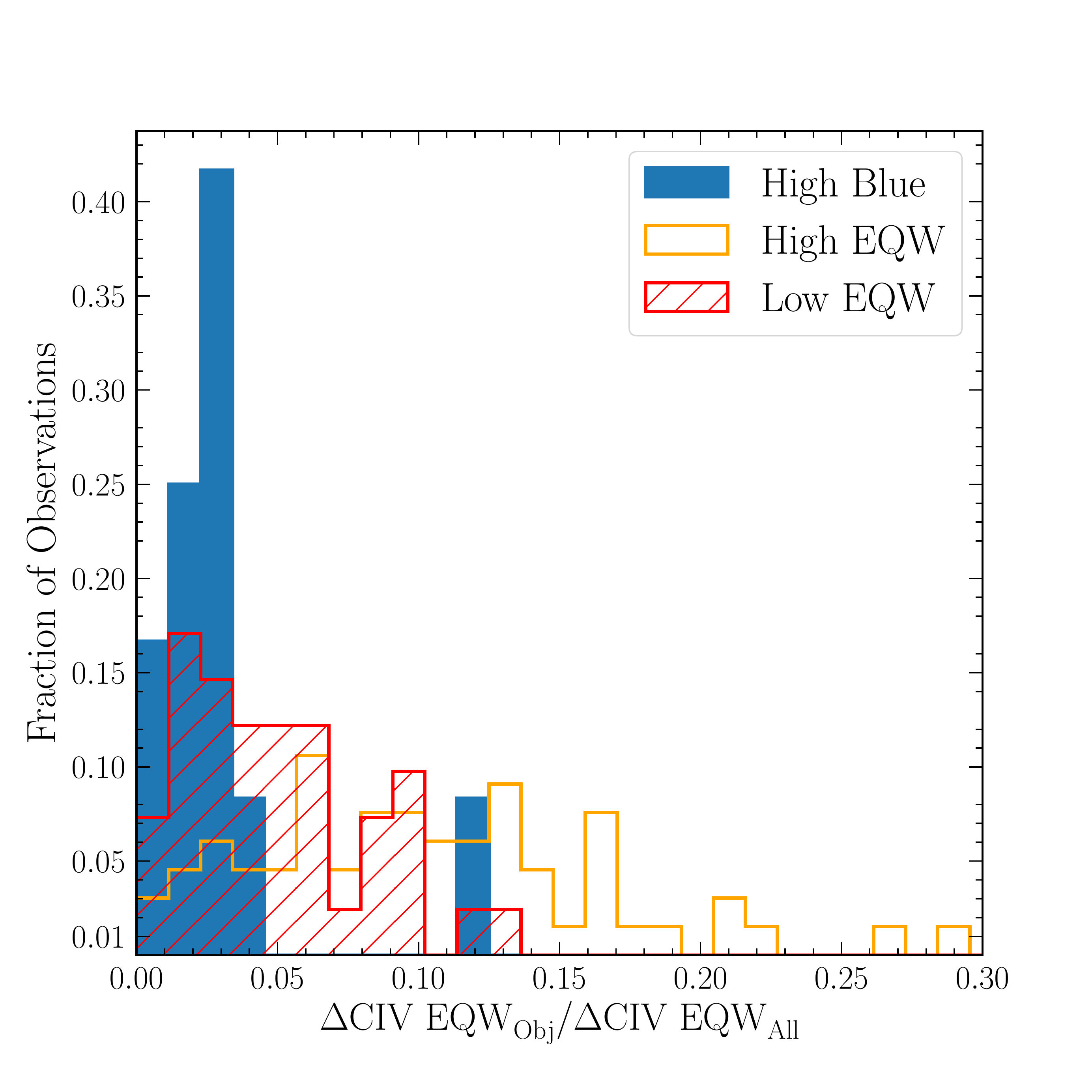}
    \caption{Histogram of the maximum change of an object's EQW over the difference between the maximum and minimum EQWs within our sample. Red: low blueshift/low EQW quasars, orange: low blueshift/high EQW quasars, and blue: high blueshift/low EQW quasars. The high blueshift quasars are much less variable in EQW than the rest of the sample.   
    \label{fig:fig14}}
\end{figure}

While Figure~\ref{fig:fig13} illustrates how much the EQW and blueshift are changing, Figure~\ref{fig:fig14} (along with Fig.~\ref{fig:fig12}) reveals {\em where} in \civ\ parameter space those changes are strongest.
\citet{Sun+2018} found little difference in the variability of the blueshift going from high EQW/low blueshift, to low EQW/low blueshift, to low EQW/high blueshift, but that low EQW/high blueshift quasars are less variable than a control sample (even after controlling for EQW, redshift, and luminosity). In Figure~\ref{fig:fig14}, we reproduce that same basic result, while Figure~\ref{fig:fig12} further reveals that high-EQW quasars are less variable in blueshift.  The highest blueshift objects will be discussed further in \S~\ref{sec:highbshift}.   Overall, these results strongly suggest that there are at least two components to the \civ\ BELR and that, in the extrema of \civ\ parameter space, one of those components completely dominates the other---consistent with the differences seen between the extrema in the \citet{Giustini+2019} model.  

\subsection{Global Baldwin Effect} 
\label{sec:beff}

\subsubsection{Variability and the Baldwin Effect}

The last of the four items mentioned at the beginning of Section~\ref{sec:varintro}, the Baldwin Effect, is an anti-correlation between the continuum luminosity and the emission-line EQW, expressed as EQW $\propto$ $L^{\beta}$ and first identified in \citet{Baldwin1977}. The Baldwin Effect holds true for many emission lines, though the slope, $\beta$, varies with the strength of the line's ionization energy \citep{Dietrich+2002} and  depends on the shape of the SED and, as a result, the choice of comparison luminosity \citep{Green1996}.  It has been additionally observed that  quasars exhibit an intrinsic Baldwin Effect, by which the EQW changes in response to an object's luminosity, generally with a slope that is much steeper (larger change in EQW with luminosity) than the global Baldwin Effect \citep{Kinney+1990}. Although we cannot investigate the intrinsic Baldwin Effect without known and accurate time lags (see Sections~\ref{sec:sample} and \ref{sec:eqw}), we can examine the global Baldwin Effect for our sample and discuss the extent to which variability (both intrinsic and driven by the continuum) can affect the calculated slope.

\begin{figure}[h!]
\includegraphics[width=3.5in]{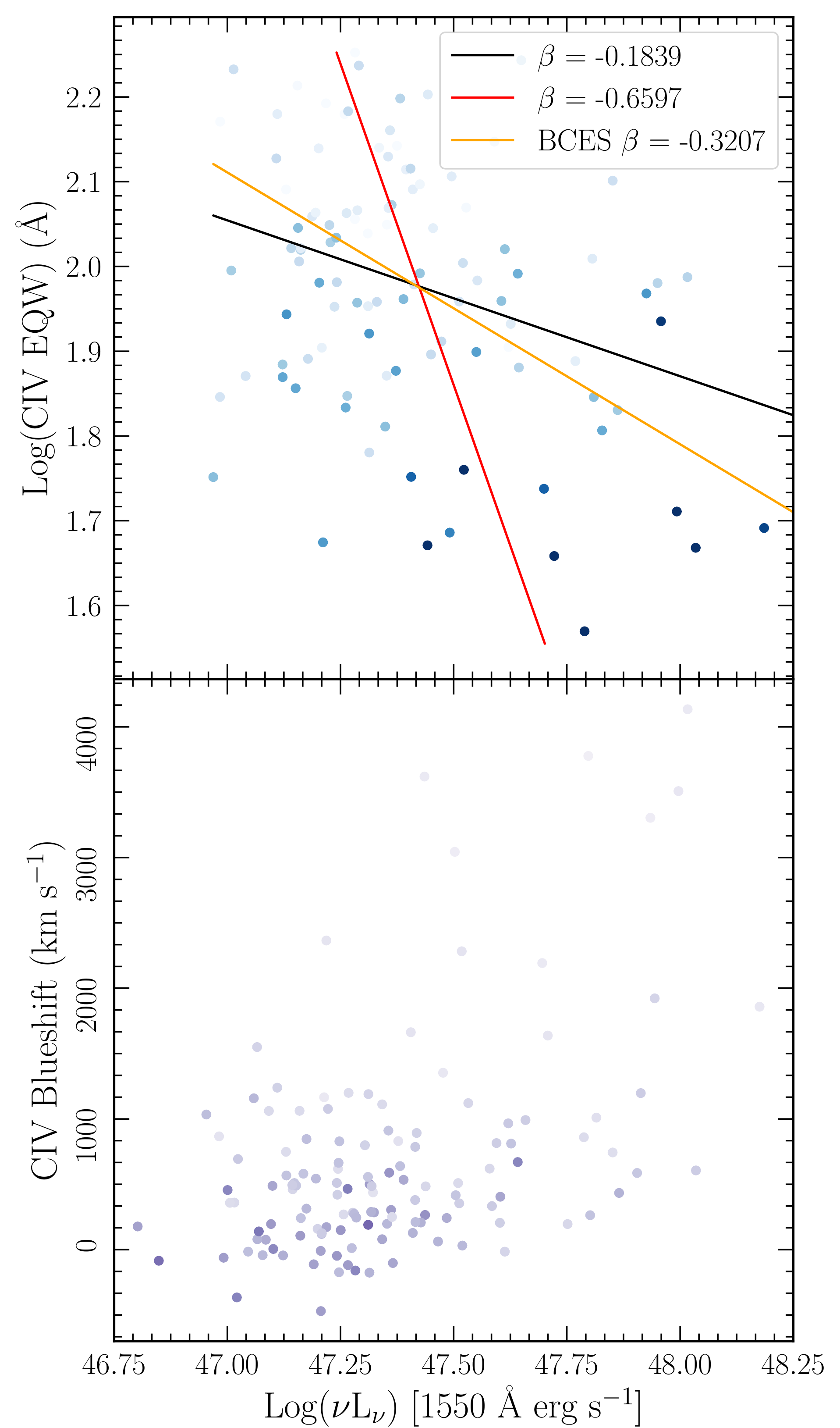}
\caption{Top: Median Log(\civ\ EQW) vs.\ median Log($\nu L_{\nu}$) at 1550\,\AA\ for 106 quasars within our sample (blue points). The points are colored by the median blueshift, with the darker shades indicating higher blueshifts. The median fit for the global Baldwin Effect for these objects is shown in black.  We additionally show a fit where EQW is instead the depedent variable (solid red line) and the BCES orthogonal of the two fits (solid orange line). 
Variability appears to have relatively little effect on the derived slope.  
Bottom: \civ\ blueshift vs.\ Log($\nu L_{\nu}$), colored by the median \civ\ EQW (darker shades indicate higher EQWs).  The increase in blueshift with luminosity is consistent with the expectation of a radiation line driven wind.
\label{fig:fig15}}
\end{figure}

Historically the $\beta$ value has been found to be $\sim-0.2$ (\citealt{Dietrich+2002}, \citealt{Baskin+2004}, \citealt{Wilhite+2006}); however, other investigations have found steeper values of $\beta$ \citep{Bian+2012,Jensen+2016}.
The top panel of Figure~\ref{fig:fig15} demonstrates the Baldwin Effect in our sample, showing the median fit in black. The points in the plot are the median values across all epochs of EQW and luminosity, colored by the median \civ\ blueshift (darker shades representing larger blueshifts). The full sample of 133 objects is not included, as not all epochs for all quasars were retained due to our S/N cuts.  For this analysis we retained the maximum number of objects with a significant number of epochs in common. This left us with 106 objects with 17 epochs of data each. We find the median $\beta$ value to be $-$0.1832, with the highest and lowest values (measured from different epochs) being $-$0.1705 and $-$0.2039, respectively. Thus variability has relatively little effect on the results.  However, fitting EQW as a function of $\nu$L$\nu$ (red solid line) or using the bivariate correlated errors and intrinsic scatter (BCES; \citealt{Akritas+1996}) orthogonal (orange) result in very different fits. The BCES orthogonal should be used when it is not clear which is the dependent variable. It calculates a line that minimizes the orthogonal distances from the points. \citet{Bian+2012} similarly reported both the historical Baldwin Effect slope and the BCES bisector slope for their samples, however, it is better to use the BCES orthogonal, as the BCES bisector can introduce error \citep{Hogg+10}. The BCES orthogonal slope, $\beta$=$-$0.3207, is consistent with the Baldwin Effect slope of quasars with mean z=2.25-2.46 as reported in \citet{Jensen+2016}.

The range of slopes we obtain for the traditional Baldwin Effect fit is consistent with the range of error in the existing literature ($\sigma_{\beta}$ $\sim$ $-$0.02 to $-$0.03). That is, if we treat each epoch as an independent experiment from which to measure the slope of the Baldwin Effect, we find that continuum variability has relatively little effect on the result (which means that correcting for time lags is relatively unimportant for investigating the global Baldwin Effect, due to the large intrinsic scatter in the relation).

\subsubsection{The Baldwin Effect and Cosmology}

One of the reasons why the Baldwin Effect has garnered interest over the years is because it was believed to be of possible use to cosmology \citep[e.g.,][]{Baldwin1977,Kinney+1990,Pogge+1992}, in part because of the very steep slope of $\beta \sim -0.64$ that was initially found, indicating a potentially sensitive mapping from EQW to luminosity. 

Unfortunately, with the much flatter slope measured from larger samples and the amount of scatter for individual objects around the relation, cosmological investigations using the Baldwin Effect have not been feasible.  A number of authors have attempted to ``tighten" the distribution by considering other parameters such as the narrow line region, [\ion{O}{3}] EQW, and $\alpha_{\rm ox}$ and marginalizing over them (respectively, \citealt{Bachev+2004,Baskin+2004,Wu+2009}) with some limited success.   In that light we colored the median EQW and $\nu$L$\nu$ values by the median blueshift in the top panel of Figure~\ref{fig:fig15} in an effort to determine if the blueshift information could be used to further constrain the relationship between log(EQW) and log($\nu$$L_{\nu}$).  However, it would seem that marginalizing over blueshift would weaken the trend, making the Baldwin Effect less useful for cosmology.

The model of  \citet{Giustini+2019} illustrates a likely source of the problem: the amount of \civ\ emission from both the wind and the failed wind depends on both \lledd\ and BH mass.  Their Figure~4 suggests that one expects similar BELR properties from both quasars with moderate \lledd\ and high BH mass as from high \lledd\ and low BH mass.
However, the X-ray properties of these two categories are very different, suggesting that there may be a way to rehabilitate the method.  Indeed, \citet{LR17} have used the correlation between UV and X-ray luminosity as a standard candle with some success.  It may be that combining the methods would yield even better results.

\subsection{Accretion Disk Winds}
\label{sec:DiskWinds}

In Section~\ref{sec:beff}, we illustrated the 
trend of EQW with luminosity.   Given the broad anticorrelation between EQW and blueshift, it is natural to expect that \civ\ blueshifts will increase with luminosity as is seen in the bottom panel of Figure~\ref{fig:fig15}.  
It is important to ask what happens to the mean blueshift at even lower luminosity (and thus lower accretion rate) than explored herein.  If the mean blueshift is consistently zero at low luminosity, that could be a sign that accretion disk winds have a luminosity/accretion-rate threshold.  On the other hand, if there exist quasars with large blueshift (measured in a way that is robust and consistent, for example not influenced by absorption), then that may indicate that winds are operating at even low luminosity (or that at least some blueshifts may be unrelated to radiation line driving).  More surprising would be to find that lower-luminosity quasars have negative blueshifts that are perhaps indicative of inflows instead of outflows.

If our analysis extended to lower luminosity, then following \citet{Veilleux+2013} and \citet{Zakamska+2014}, we might expect to find that below a threshold of $L_{\rm Bol} \sim 3\times10^{45}$ ergs/s (well below that shown in Fig.~\ref{fig:fig15}), there is a lack of objects with significant blueshifts.  Such a threshold would be consistent with a minimum luminosity being a necessary, but insufficient condition for radiation line driven accretion disk winds.
    
\subsection{High-blueshift and Weak-lined Quasars}
\label{sec:highbshift}

The analysis by \citet{Sun+2018} suggested that high-blueshift, low-EQW quasars are less variable in terms of their \civ\ blueshifts, while \citet{Meusinger+11} and Moreno et al. (2020, in preparation) find that so-called ``weak-line quasars" (WLQs) are less variable than other quasars.  Our analysis enables us to try to connect these results and make predictions about such sources.

We define a high-blueshift subsample as quasars with median \civ\ blueshift $\ge$ 1500 km\,s$^{-1}$. While these sources also have the smallest EQWs in the sample, they are not nearly as weak in \civ\ as traditionally defined WLQs (which have 
EQW $<$ 15 $\textrm{\AA}$; see, e.g.,  \citealt{DS+2009,Plotkin+2015}), but they do possess relatively small EQWs, a high \lledd, and may be expected to have similar geometries to this population.  Thus, our high-blueshift, low-EQW sample can provide a sort of ``bridge" between the most extreme quasars and more normal quasars, much as the sample from \citet{Ni+2018}.  

For quasars with an \lledd\ $\ge$ 0.3, the accretion disk is not likely to conform to the standard thin accretion disk model \citep{Jiang+2014,Luo+2015}. A high accretion rate in a quasar is thought to result in a ``slim" inner disk which prevents the ionizing continuum from exciting the BELR gas (\citealt{Abramowicz1988}, also see Figure 18 from \citealt{Luo+2015}). The presence of an inner slim accretion disc would aid in the launching of a radiation driven disk wind, causing larger blueshifts \citep{Murray+1995,Proga+2000}. 

Thus, it is interesting that we and other authors find quasars at this extreme to be less variable than more typical quasars \citep{Wilhite+08,Meusinger+11}.  
From our Figures~\ref{fig:fig12} and \ref{fig:fig14}, it can be seen that the high blueshift quasars are less variable not only in their blueshifts\footnote[5]{The exceptions to this trend are mostly due to low S/N and absorption features present in the spectra. One has asymmetric Component 3, and therefore has real variability in blueshift.}, they also are less variable in EQW.  This lack of EQW variability indicates that neither the continuum, nor the BELR is changing significantly for these quasars (as even if the BELR is not varying, changes in the continuum will cause the EQW to change as discussed above). In simulations of AGN with slim disks, \citet{Jiang+2019} find that the accretion disk is supported by magnetic rather than thermal pressure. Additionally, the surface density of the accretion disk increased with increasing accretion rate, and Moreno et al. (2020, in preparation) suggests that this increase in density reduces the effects of small-scale perturbations in the disk, resulting in the lower variability of WLQs.

\subsection{Associated Absorption Lines}
\label{sec:AAL0s}

Narrow absorption lines in quasar spectra come in three ``flavors": those that are due to intervening galaxies, those that are due to high-velocity outflows, and those that are associated with the host and are at or near the systemic redshift \citep{Weymann+1979, Stone+2019}.  Associated absorption lines (AALs) include the latter two types where the line is within $\pm 1000$ km $\textrm{s}^{-1}$ of the systemic redshift of the quasar \citep{Foltz+1986,Hewett+2010}.  It is well known that there is an excess of AALs at the systemic redshift (hereafter AAL0s) in steep-spectrum radio-loud quasars (particularly lower luminosity sources), but \citet{Stone+2019} showed that AAL0s are just as common in radio-quiet sources (their Figures~4a and 9).  Given that steep-spectrum radio-loud sources are thought to be more edge-on systems, \citet{Stone+2019} suggested that AAL0 systems could be an orientation indicator for radio-quiet quasars.

That galaxies are reservoirs of partially ionized gas is well known to quasar astronomers as the absorption features due to galaxies litter quasar spectra.  Thus, if anything, it is surprising not to see systemic \civ\ along all lines of sight.  That it is rare suggests that quasars are capable of clearing out this gas.  Where it remains is therefore much more likely to be a location protected from winds, such as when the disk is more edge on to our line of sight.
This idea is particularly interesting for our investigation since the SDSS-RM spectra have sufficient S/N to detect even relatively weak AAL0s.  Given that \citet{Shen+2014} argue that the FWHM of H$\beta$ may have a significant contribution from orientation and not just from BH mass, it is important to understand if the \civ\ emission line also exhibits evidence of an orientation dependence.  

\begin{figure}[h!]
\includegraphics[width=3.5in]{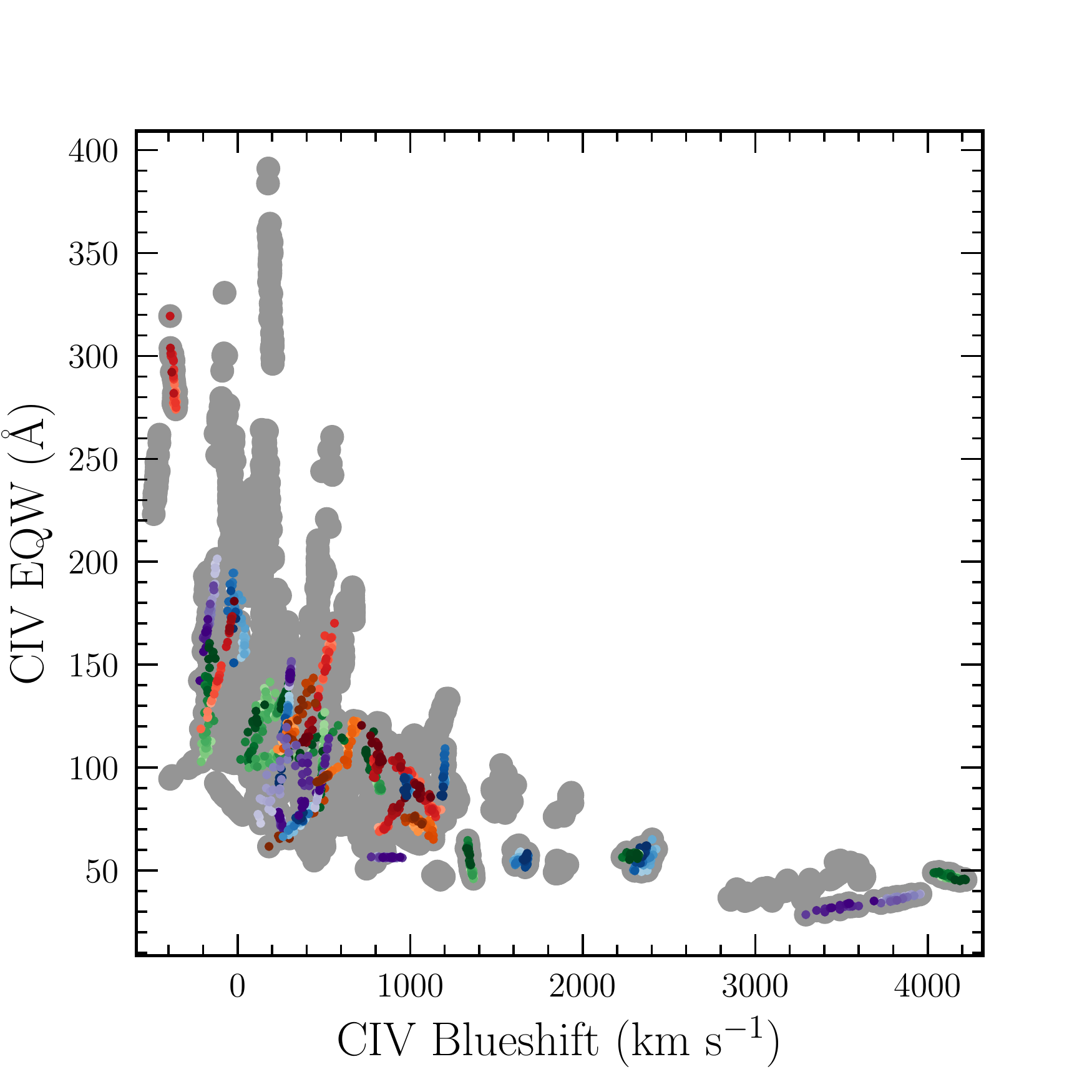}
\caption{The \civ\ parameter space with the AAL0s highlighted. Grey points show the full distribution of CIV parameter values from our sample. If AAL0 systems indicate more edge-on orientation, then the lack of a clear bias of such sources suggests that there may not be an orientation dependence to this parameter space. 
\label{fig:fig:fig16}}
\end{figure}

\begin{figure*}[th!]
    \includegraphics[width=7in, trim= 1cm 0 0 0]{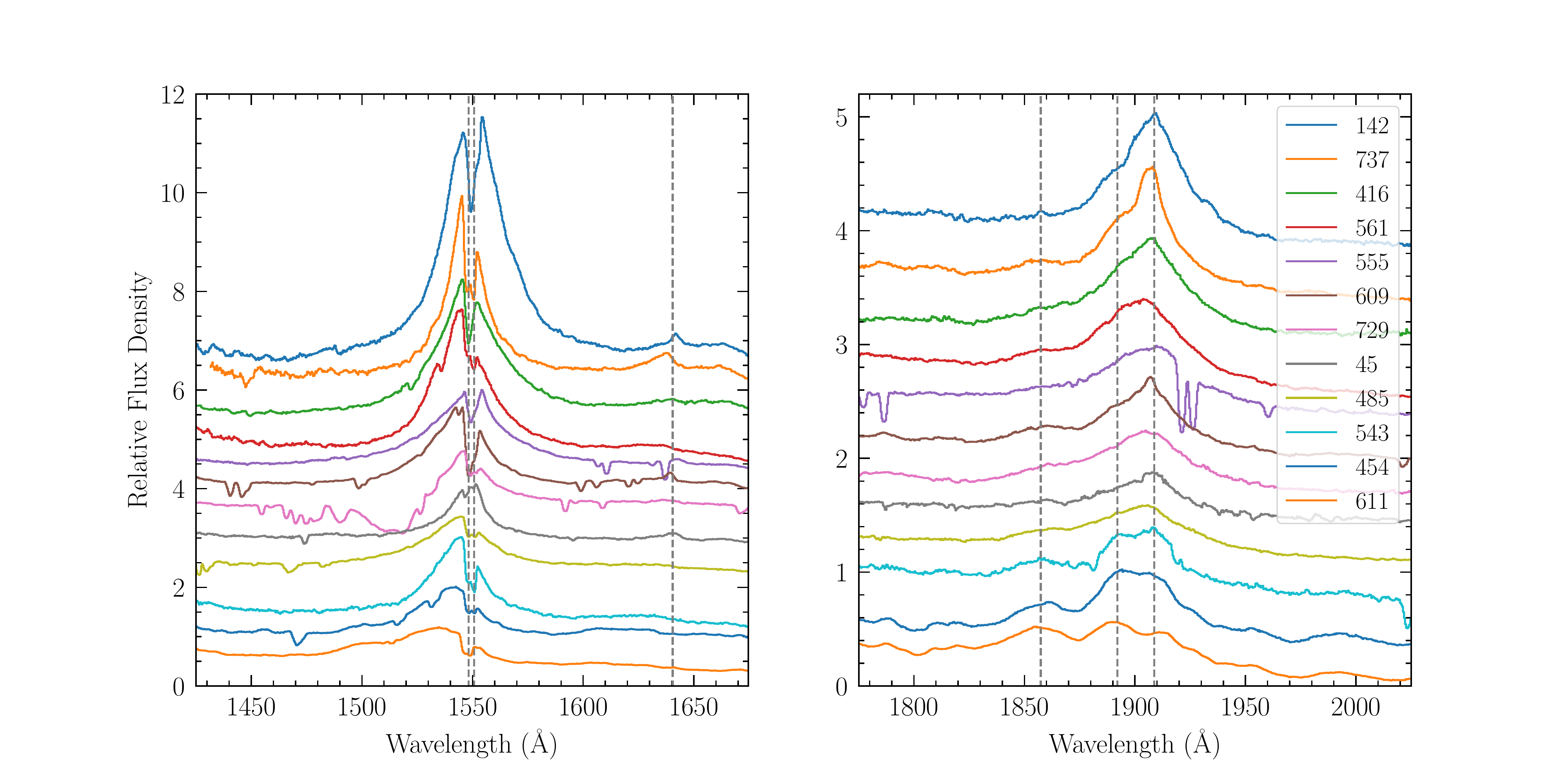}
    \caption{The \civ\ line (left) and \ion{C}{3}] line complex for 12 quasars exhibiting AAL0s in our sample. The dashed vertical lines are placed at the laboratory wavelengths of \civ, \ion{He}{2}, \ion{Al}{3}, \ion{Si}{3}], and \ion{C}{3}].  These objects span the full range of \civ\ space, which is illustrated not only by the evolution of the \civ\ blueshift and EQW from top to bottom, but also the ratios of \ion{C}{3}] and \ion{Al}{3} to \ion{Si}{3}], which are strongly correlated with \civ.
    \label{fig:fig17}}
\end{figure*}

Figure~\ref{fig:fig:fig16} shows the distribution of objects that we flag as AAL0s within the \civ\ parameter space, while
Figure~\ref{fig:fig17} shows the spectra of 12 example AAL0 quasars from our sample from high EQW/low blueshift (top) to low EQW/high blueshift (bottom).  These figures emphasize how the AAL0s span the full range of \civ\ parameter space from high EQW/low blueshift to low EQW/high blueshift.  In Figure~\ref{fig:fig17} many of the absorption features are redward of the line peak, illustrating that the peak of \civ\ emission can be a poor indication of the systemic redshift of quasars \citep{Hewett+2010}, which is better indicated here by the \ciii\ line complex. We have included only those quasars for which the \civ\ doublet is resolved and there is no obvious sign of strong \ion{Si}{4} absorption (which appears to distinguish these systems from miniBALs).  The fact that these AAL0s span the full range of \civ\ space and that there is no systematic bias within that trend towards higher/lower EQW or blueshift suggests that orientation is not a primary factor in the location of quasars in \civ\ space (assuming that the AAL0s indeed are indicative of more edge-on orientation).

We find that 10--20\% of the objects in our SDSS-RM subsample exhibit AAL0s.  Thus if the presence of an AAL0 system is indeed indicative of more edge-on orientation, such objects could enable statistical investigations surpassing those using radio spectral index as an orientation indication (as the fraction of radio-detected quasars is smaller than 10\% and those with robust radio spectral indices are even more rare; e.g., \citealt{KR2015}).
Our finding leads to a novel prediction that would help to determine if the edge-on hypothesis is valid. \citet{Luo+2015} and \citet{Ni+2018} have argued that WLQs can be both X-ray normal or X-ray weak. This is thought to be an orientation effect, with WLQs viewed edge-on as being observed to be X-ray weak. If the WLQ model is correct and if AAL0s are edge on, then, in the context of the model put forth by \citet{Luo+2015} and \citet{Ni+2018}, we make the following prediction: AAL0 systems seen in quasars where there is evidence for strong winds and/or a slim accretion disk (such as WLQs or very high blueshift quasars) will be X-ray weaker than similar quasars without AAL0s. However, quasars where there is no evidence for strong winds or a slim accretion disk will not differ in their X-ray properties as a function of the presence or not of AAL0s.  That is because the systems of interest are special in two ways.  First, they have a special accretion disk geometry; second, they have a special orientation with respect to that geometry.  New X-ray observations and/or archival data of a sufficiently large sample of quasars with high enough spectral S/N and resolution (sufficient to resolve the \civ\ doublet in absorption) would be able to test this prediction.

\subsection{An Empirical Quasar Main Sequence}

At the beginning of this paper we discussed both the physical parameters that control the appearance of quasars: luminosity, accretion rate, black hole mass, orientation, and spin and the need for a ``main sequence" for quasars \citep[e.g.,][]{Marziani+2018}.  We further discussed the changes expected in the emission lines in the context of the multi-component (wind and failed wind) model of \citet{Giustini+2019}.  While \citet{Marziani+2018} make a case for a quasar main sequence founded in properties measured from quasar spectra at low redshift (from which is it possible to determine relatively robust BH masses from the H$\beta$ emission line), it is well-known that BH masses at high redshift derived from \civ\ are much less robust \citep[e.g.,][]{Coatman+2017}.  Coupled with uncertainty about the bolometric correction for a diverse sample of quasars \citep{Krawczyk+2013}, we are potentially left without accurate estimates of {\em any} of the physical parameters needed to construct a main sequence at high redshift.  

Nevertheless, there is evidence that the \civ\ empirical parameter space traces trends in physical parameter space, especially in terms of identifying extrema \citep[e.g.,][]{Sulentic+2007,Richards+2011,Rankine+2020}.  As such it is important to realize that, even without the parameters that are needed to construct a ``physical" main sequence for quasars, we have the information that we need to construct an ``empirical" main sequence.  Furthermore, that single-epoch spectra, even in the absence of knowledge of time lags can be used to define such an empirical sequence, using models like \citet{Giustini+2019} to provide a physical framework.  

Future refinement of the ICA technique to enable a single set of components to reconstruct quasars spanning the full diversity of quasars may then enable such a main sequence to be described by just a few component weights that capture essentially all of the information in the UV part of quasar spectra.  Indeed, such a system could then be used to guide the very research needed to turn UV spectra into the physical parameters that are needed to construct an actual physical main sequence.

\section{Conclusions}
\label{sec:conclusion}

We have used the ICA technique to analyze a sample of 133 SDSS quasars with time-resolved spectroscopy over $\sim$ 900 days from the SDSS-RM project.  Our primary findings are (1) ICA reconstruction allows for more robust characterization of the line-emitting gas, essentially by smoothing over noisy data. (Sections \ref{sec:individualica}, \ref{sec:CombRecon} and Figures \ref{fig:fig7} and \ref{fig:fig8}); (2) Only two ICA components are needed to reconstruct most of the spectra at high fidelity, whereas three components are expected to be needed in cases where reverberation mapping analysis is likely to be successful (Section \ref{sec:individualica}, Figure \ref{fig:fig6}); (3) Single-epoch spectroscopy exaggerates the apparent variability of EQW due to inherent dependence on continuum, which cannot be easily corrected for the long lags expected for luminous quasars (making it impossible to investigate the intrinsic Baldwin Effect) (Section \ref{sec:lineparams}); (4) Multi-epoch spectroscopy gives a more robust indication of the variability of the line strength and reveals that quasars do not significantly change location in terms of their \civ\ emission-line parameters (Figure~\ref{fig:fig14}) over the timescales investigated and single-epoch results do not have a significant impact on investigations of the global Baldwin Effect. (Section \ref{sec:beff}, Figure \ref{fig:fig15}); 
(5) Quasars with emission line properties indicative of higher \lledd\ are less variable (both in blueshift and EQW), consistent with models with enhanced accretion disk density. (Section \ref{sec:highbshift}, Figures \ref{fig:fig12} and \ref{fig:fig14}); and (6) Narrow absorption features at the systemic redshift may be indicative of orientation (including radio-quiet quasars) and may appear in as much as 20\% of the quasar sample. (Section \ref{sec:AAL0s} and Figures~\ref{fig:fig:fig16} and \ref{fig:fig17}).

In Sections~\ref{sec:beff} and Figure \ref{fig:fig15} we found that analysis of the Baldwin Effect and the luminosity dependence of blueshifts was limited by the relatively narrow range in luminosity. \citet{Dietrich+2002} found that the lack of dynamic range in luminosity can bias the calculated Baldwin Effect slope.  Thus, in future work we will probe distribution of EQWs and blueshifts to lower luminosity in a systematic way, particularly to determine if there is evidence for a luminosity dependence to the presence of radiation line driven winds.  For example by analyzing the median spectra of the lower S/N SDSS-RM spectra and also objects in the {\em Hubble Space Telescope (HST)} archive that are included in the SDSS sample and have {\em HST} coverage of \civ\ with high S/N.\\

\acknowledgements
This material is based upon work supported by the National Science Foundation under Grant No.\ AST-1908716.
Support for Programs HST-GO-14135.001-A and HST-AR-15048.001-A was provided by NASA through a grant from the Space Telescope Science Institute, which is operated by the Association of Universities for Research in Astronomy, Incorporated, under NASA contract NAS5-26555. Funding for the Sloan Digital Sky Survey IV has been provided by the Alfred P. Sloan Foundation, the U.S.\ Department of Energy Office of Science, and the Participating Institutions. SDSS-IV acknowledges support and resources from the Center for High-Performance Computing at the University of Utah. The SDSS web site is www.sdss.org. SDSS-IV is managed by the Astrophysical Research Consortium for the Participating Institutions of the SDSS Collaboration including the Brazilian Participation Group, the Carnegie Institution for Science, Carnegie Mellon University, the Chilean Participation Group, the French Participation Group, Harvard-Smithsonian Center for Astrophysics, Instituto de Astrofisica de Canarias, The Johns Hopkins University, Kavli Institute for the Physics and Mathematics of the Universe (IPMU) / University of Tokyo, the Korean Participation Group, Lawrence Berkeley National Laboratory, Leibniz Institut f{\"u}r Astrophysik Potsdam (AIP), Max-Planck-Institut f{\"u}r Astronomie (MPIA Heidelberg), Max-Planck-Institut f{\"u}r Astrophysik (MPA Garching), Max-Planck-Institut f{\"u}r Extraterrestrische Physik (MPE), National Astronomical Observatories of China, New Mexico State University, New York University, University of Notre Dame, Observatario Nacional / MCTI, The Ohio State University, Pennsylvania State University, Shanghai Astronomical Observatory, United Kingdom Participation Group, Universidad Nacional Autonoma de Mexico, University of Arizona, University of Colorado Boulder, University of Oxford, University of Portsmouth, University of Utah, University of Virginia, University of Washington, University of Wisconsin, Vanderbilt University, and Yale University.
We thank Trevor McCaffrey for help creating the best-fit curve through the \civ\ parameter space and Vivienne Wild for discussions about the AAL0 population.

\bibliography{sdssrm}

\end{document}